\documentclass[11pt,a4paper]{article}
\usepackage[english]{babel}

\usepackage[bitstream-charter]{mathdesign}
\usepackage[T1]{fontenc}

\usepackage{geometry}
\usepackage{graphicx}
\usepackage[export]{adjustbox}
\usepackage[dvipsnames]{xcolor} 
\usepackage{float}
\usepackage{wrapfig}

\usepackage{setspace}

\usepackage{url}
\usepackage{hyphenat}
\usepackage{amsmath} 
\usepackage{relsize}

\usepackage{xhfill}
\usepackage{tocloft}

\usepackage[round,sort,comma,super,authoryear]{natbib}

\usepackage{wrapfig}
\usepackage{indentfirst}
\usepackage{hhline}

\usepackage[small]{caption}

\xdefinecolor{RED}{named}{BrickRed}
\xdefinecolor{linkcolor}{named}{BlueViolet}
\definecolor{background}{RGB}{209, 222, 255}
\definecolor{background2}{RGB}{0, 40, 151}
\everymath=\expandafter{\the\everymath\displaystyle}

\usepackage[colorlinks=true,
pdfstartview=FitV,
linkcolor= linkcolor,
citecolor= linkcolor,
urlcolor= linkcolor,
hyperindex=true,
hyperfigures=false]
{hyperref}

\DeclareCaptionFont{blue}{\color{linkcolor}}
\newcommand{\mbf}{\boldsymbol}

\newcommand{\p}{\partial}

\newcommand{\ddroit}[2]{\ensuremath{\dfrac{\mathrm{d}{#1}}{\mathrm{d{#2}}}}}

\newcommand{\bu}{\mbf{u}}

\newcommand{\ie}{\textit{i.e.}~}

\usepackage{tikz}
\newcommand*\circled[1]{\tikz[baseline=(char.base)]{
            \node[shape=circle,draw,inner sep=1pt] (char) {#1};}}

\usepackage{tabularx} 

\begin{document}


\thispagestyle{empty}

\vspace{0.5cm}

\noindent \textsc{Centre International des Sciences M\'ecaniques}\\
Fluid Mechanics of Planets and Stars\\
April 16-20th 2018\\

\vspace{0.5cm}

\begin{center}

\vspace{1.5cm}

\textcolor{RED}{\rule[11pt]{\linewidth}{2pt}}
\begin{doublespacing}
\textbf{\huge{ \textsc{Rotational dynamics of planetary cores}}}\vspace*{0.25cm}\\
\Large{Instability driven by precession, libration and tides}
\end{doublespacing}
\textcolor{RED}{\rule[-11pt]{\linewidth}{2pt}}


\vspace{0.75cm}

\textbf{Thomas \textsc{Le Reun} and Michael \textsc{Le Bars}}

Aix-Marseille University, CNRS, Centrale Marseille, IRPHE UMR 7342, Marseille, France

\end{center}

\vspace{0.75cm}

\begin{center}
\textcolor{RED}{\textbf{\large{Abstract}}}  
\end{center}
In this chapter, we explore how gravitational interactions drive turbulent flows inside planetary cores and provide an interesting alternative to convection to explain dynamo action and magnetic fields around terrestrial bodies. In the first section, we introduce tidal interactions and their effects on the shape and rotation of astrophysical bodies. A method is given to derive the primary response of liquid interiors to these tidally-driven perturbations. In the second section, we detail the stability of this primary response and demonstrate that it is able to drive resonance of inertial waves. As the instability mechanism is introduced, we draw an analogy with the parametric amplification of a pendulum whose length is periodically varied. Lastly, we present recent results regarding this instability, in particular its non-linear saturation and its ability to drive dynamo action. We present how it has proved helpful to explain the magnetic field of the early Moon. 

\vspace{0.5cm}

\vspace{0.5cm}


\setlength{\parindent}{10pt}

\newpage
\thispagestyle{empty}
{ \hypersetup{linkcolor=black}
\setcounter{tocdepth}{2}
\tableofcontents
}
\thispagestyle{empty}


\newpage



\section*{Introduction: From planetary magnetic fields to core turbulence}
\addcontentsline{toc}{section}{Introduction: From planetary magnetic fields to core turbulence}

\begin{figure}
\includegraphics[width=\linewidth]{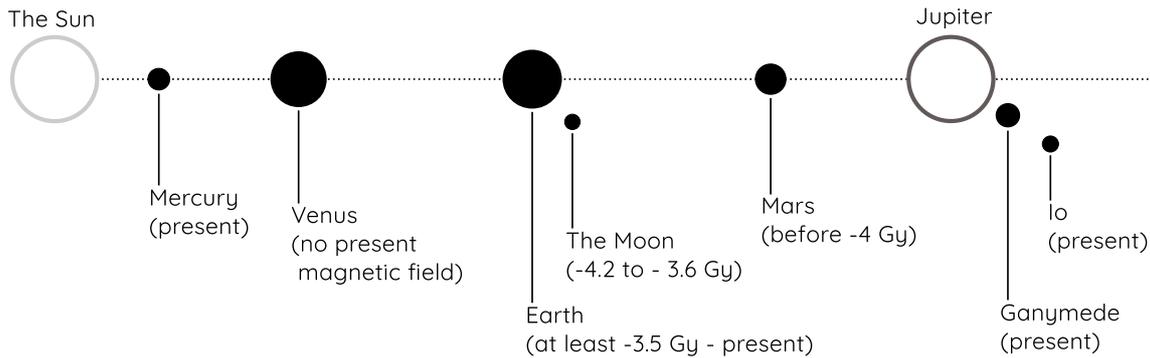}
\caption{A summary of the terrestrial bodies (in black) of the Solar System known for having a past or present dynamo. The relative size of these planets and moons is respected (apart for Jupiter and the Sun). Venus is given for comparison as it is not surrounded by a magnetic field although it is of similar size to the Earth. It is not known whether Venus had a magnetic field in the past. We report the estimated period of existence of these fields when it is known, based on \cite{garrick-bethell_early_2009,tarduno_geodynamo_2010,stevenson_mars_2001}. }
\label{fig:terrestrial_planets}
\end{figure}

In addition to the Earth, several terrestrial bodies of the Solar System are known to be presently protected from solar radiation by an intense magnetic field, or present evidence of a past one. 
For instance, flybys operated by probes equipped with magnetometers have revealed the presence of a magnetic field surrounding Mercury as well as Ganymede and Io, two of Jupiter's largest moons \citep{ness_magnetic_1975,kivelson_discovery_1996,
sarson_magnetoconvection_1997,showman_galilean_1999,
kivelson_permanent_2002}. 
Magnetized rock samples from Mars and the Earth's Moon have also revealed the existence of a past intense magnetic field \citep{stevenson_mars_2001,garrick-bethell_early_2009,le_bars_impact-driven_2011}. 
A summary of what is presently known about magnetic fields of terrestrial planets in the Solar System is given in figure \ref{fig:terrestrial_planets}. 
Beyond the Solar System, magnetic fields are also expected in extra-solar planets, where they constitute one of the key ingredients for habitability.

As first conjectured by \cite{larmor_how_1919}, the magnetic field of a planet originates in the turbulent motion of liquid iron in its core \citep{olson_8.01_2015}. 
Following the seminal works of \cite{roberts_thermal_1968} and \cite{busse_thermal_1970}, it has been shown that buoyancy-driven flows such as thermal and solutal convection in cores provoke turbulent stirring and dynamo action \citep{glatzmaiers_three-dimensional_1995}.
This convective motion is driven by the secular cooling of a planet, by radiogenic heating, and by latent heat and potential energy release during its core solidification.
While the energy budget to sustain the present day magnetic field of the Earth is closed ---even if still partly controversial, see e.g. \cite{labrosse_thermal_2015}---, the Earth early dynamo and the dynamo in smaller bodies remain largely unexplained. 
As an illustration, let us assume that the main source for dynamo lies in the initial thermal energy of the body acquired during its formation. 
This initial thermal energy can be estimated assuming that it is roughly tantamount to the loss of gravitational potential energy from a dispersed cloud to an aggregated body, and is proportional to $R^5$ where $R$ is the radius of the planet at the end of its formation. 
As the secular heat loss scales like the surface of a planet, we can infer that a typical planetary cooling time grows like $R^3$.
As a consequence, the core of relatively small planetary bodies such as Ganymede, Mercury or the early Moon cools down very quickly and should not be able to sustain turbulent convective motion and long-term dynamo action. 
Even for larger planets such as the Earth, the initial temperature to maintain a dynamo all along their lifetime should be extremely hot, in possible contradiction with the presence of a solid mantle at the beginning of their existence \citep{andrault2016deep}.

However, initial heat is not the only source of energy available to drive fluid motion. 
In particular, a huge amount of mechanical energy is stored in the rotational dynamics of planetary systems \citep{le_bars_flows_2015}. 
If it is possible to convert this mechanical energy into turbulent kinetic energy inside a planetary core, it provides an interesting alternative to convective instabilities to drive planetary dynamos.
In the ideal case of a perfectly spherical planet with uniform rotation, this conversion cannot happen: the liquid core follows the terrestrial planet in its solid-body rotation. 
However, tidal interactions between astrophysical bodies result in periodic alteration of their shape, of the direction of their rotation axis and of their rotation rate, which can then force fluid motion inside their cores. 

This idea that tidal interactions force core turbulence was first introduced by Malkus in three seminal articles \citep{malkus_precessional_1963,malkus_precession_1968,malkus_experimental_1989}, but was largely dismissed by geophysicists for decades. 
As noted by {Kerswell}, this was mainly due to a misunderstanding regarding the nature of the flow excited by tides 
\citep{kerswell_upper_1996,kerswell_elliptical_2002}. 
Tidal interactions are of small amplitudes and their direct forcing only generates small departures from the solid-body rotation of the fluid core. 
Alone, these small perturbations are not powerful enough to sustain any magnetic field. 
However, these periodic alterations are able to excite resonant instabilities which can then break down into bulk-filling turbulence.

%
%
While the flow directly created by tidal perturbations is purely laminar and low amplitude in the first place, the excited resonant instabilities are responsible for converting the huge rotational kinetic energy into turbulence, and possibly dynamo action. 
Flows driven by tidal instabilities in a geophysical context have benefited from extensive investigation over the past two decades.
Theoretical  and experimental studies have revealed that these instabilities, for the most part, rely on the interplay between inertial waves ---which exist in any rotating fluid because of the restoring action of the Coriolis force--- and the harmonic forcing.
The underlying mechanism is a sub-harmonic resonance called the elliptical instability
\citep{kerswell_instability_1993,le_dizes_three-dimensional_2000,lacaze_elliptical_2005,le_bars_coriolis_2007,cebron_libration-driven_2014,grannan_experimental_2014,lemasquerier_d._librationdriven_2017}.
In particular, this research has clarified the conditions for such an instability to develop in terms of tidal forcing versus viscous damping inside planetary cores \cite[see for instance][]{cebron_elliptical_2012}. 
Yet there is still much to understand about tidally-driven instabilities; in particular, the comprehension of their saturation into bulk-filling turbulence remains a challenging problem although of crucial importance to predict the resulting dynamo action. 
Such a turbulence bears many particularities compared to classical turbulence. 
In addition to being strongly influenced by rotation, this turbulence is forced at low amplitude ---tidal perturbations are weak forcings--- in a very small dissipation regime ---because of the massive size of the considered bodies. 
This problem is thus difficult to study as such regimes are far beyond the reach of any numerical simulation or laboratory experiment.
Only careful extrapolations to planetary cores can be drawn from present knowledge.
Nevertheless, significant steps have been made over the past few years \cite[see for instance][]{favier_generation_2015,grannan_tidally_2017,le_reun_inertial_2017}, one of the most striking results being the evidence of a fully turbulent kinematic dynamo driven by tidal forcing in a planetary-relevant ellipsoidal geometry \citep{reddy_k._sandeep_turbulent_2018}. 

In this chapter, we aim at providing a simple understanding of tidally-driven instabilities.
In the first section, we dwell on the primary response of planetary cores to tidal forcings.
The second section is devoted to the parametric instabilities which develop on this primary response; in particular, we draw an analogy with the resonance of length-varying pendulums and exhibit the mathematical underpinnings of tidally-driven instabilities.
In the last section, we review a few recent results and challenges brought by the study of mechanical forcings in planetary cores.

\section{Tidal forcings in planetary cores: the primary response to tides}
\vspace*{1cm}

The aim of this section is to review the basic effects of tides on planetary cores. 
We introduce the principal perturbations to the rotational motion of planets induced by tides and present the method to infer the primary response of a fluid cavity to those perturbations.

\subsection{The shape of a planet undergoing tidal distortion}

%
%
\begin{figure}
\centering
\includegraphics[width=0.8\linewidth]{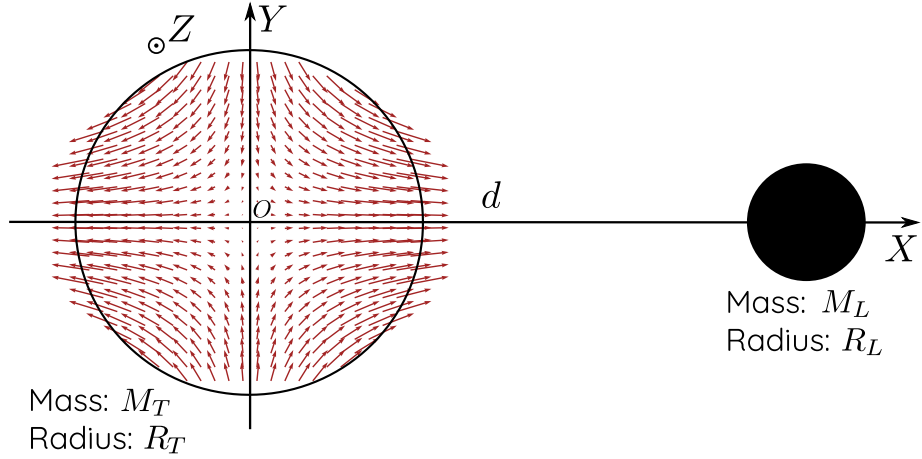}
\caption{Schematic diagram of two astrophysical bodies such as the Earth and the Moon, assuming that they are perfectly spherical and homogeneous. The tidal force field induced by the Moon is represented by the red arrows in the Earth. This field is invariant under rotation around the $X$ axis. Note that the Earth induces a similar field inside the Moon.}
\label{fig:tidal_scheme_static}
\end{figure}
As seen in another chapter of this book, the gravitational interaction between two astrophysical bodies results in a force field distorting them called ``tides''. 
For each body, it reflects the difference between the total gravitational attraction which drives its motion and the local attraction. 
Considering a planet $T$ and a moon $L$ separated by a distance $d$ taken constant in first approximation ---see figure \ref{fig:tidal_scheme_static}---, the tidal potential writes at lowest order:
\begin{equation}
\label{eq:tidal_potential}
U_{\mathrm{tides}} ~=~ \frac{G M_T}{R_T^3} ~ \frac{M_L}{M_T} \left( \frac{R_T}{d} \right)^3~ \left( r^2 - (\mbf{e}_X \cdot \mbf{r} )^2 \right)
\end{equation}
with $r^2 = (X^2+Y^2 + Z^2) $, $ \mbf{r}= X \mbf{e}_X + Y \mbf{e}_Y + Z \mbf{e}_Z $ and $G$ the gravitational constant ---the remaining variables are defined on figure \ref{fig:tidal_scheme_static}. 
The tidal force field is represented in figure \ref{fig:tidal_scheme_static} and bears two important symmetries: it is invariant by rotation around the planet-moon axis $(OX)$ and by reflection relative to the $(YOZ)$ plane. 
The deformation induced by the tidal potential (\ref{eq:tidal_potential}) can be analytically determined (see for instance discussions in \cite{barker_nonlinear_2016} and \cite{barker_nonlinear_2016-1}). 
At the lowest order, it can be shown that the planet adopts an ellipsoidal shape; the combination of both equatorial ---or rotational--- and tidal bulges forces the three axes of this ellipsoid to have different lengths. 
In the following, we assume that the outer boundaries of planetary cores are ellipsoidal. 
For simplicity's sake, we do not take into account the possible presence of a solid inner core. In a general context, as the rigidities of a solid iron inner core and of the rocky mantle are different, the liquid iron domain would be a shell confined between two ellipsoids which are not necessarily homothetic.
Such a geometry does not change the overall dynamics but makes its analysis more difficult ---see for instance \cite{lemasquerier_d._librationdriven_2017}. 
In addition, the case we consider is an actual situation encountered in young planets as solid inner cores only crystallize later after their formation. 
For instance, the Earth is known to have been surrounded by a magnetic field since at least 3.5 Gy although the inner core is only around 1 Gy old or less \citep{labrosse_thermal_2015}. 

\subsection{Flow driven by differential spin and orbit}
\label{tidal_flow}

In a configuration similar to the Earth-Moon system, the spin of the Earth is not synchronized with the orbit of the Moon. Because the Earth rotates every day, but the Moon's orbit is 27 days long, and because of the symmetry of the tidal field, the solid part of the Earth is subject to slightly less than two tidal rises per day. 
Focusing on the core, while the liquid iron rotates at the same rate as the Earth, its outer shape bears a tidal bulge which follows the orbiting motion of the Moon.

\begin{figure}
\centering
\includegraphics[width=0.55\linewidth]{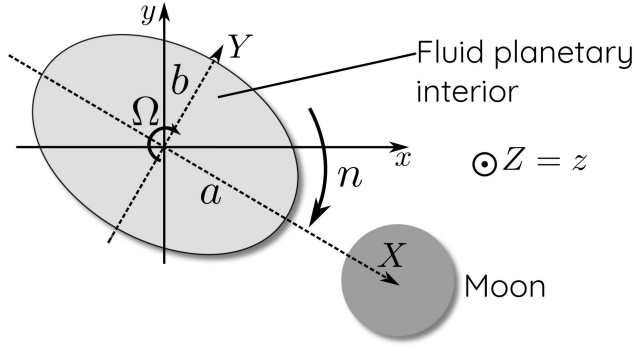}
\caption{Schematic diagram of a moon orbiting around a planet. We define two system of axes: $(OXYZ)$ follows the revolution of the moon (hence of the ellipsoidal shape of the planet) and rotates at rate $n \mbf{e}_z$; $(Oxyz)$ tracks the rotation of the solid mantle of the plane and rotates at rate $\Omega \mbf{e}_z$. We assume that the two rotation motions occur in the same plane for simplicity's sake. The fluid envelop is ellipsoidal, its axes have length $a$ and $b$ in the equatorial plane, and $c$ along the axis of rotation.}
\label{fig:tidal_flow_scheme}
\end{figure}

\begin{figure}
\centering
\includegraphics[width=0.8\linewidth]{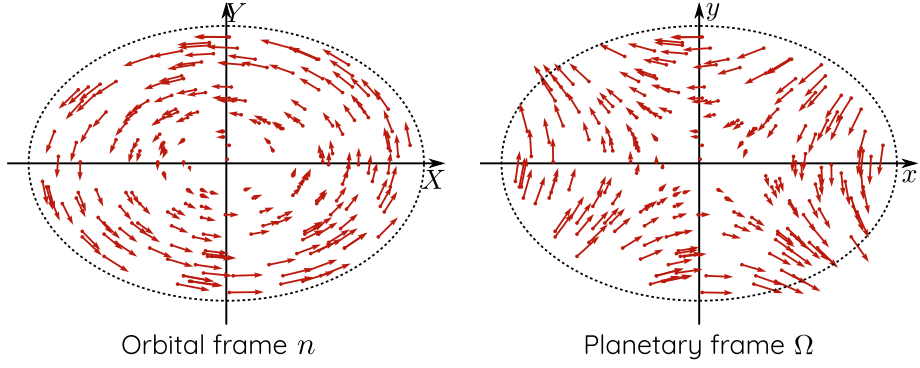}
\caption{Tidal flow velocity field seen from the orbital frame rotating at rate $n$ (left) and from the planetary frame rotating at rate $\Omega$ (right). The arrows scale is not the same on the two figures. $ \Omega-n$ is positive: the fluid moves counter clockwise in the orbital frame and the elliptical bulge moves clockwise when seen from the planetary frame. Planetary frame highlights the strain field which perturbs the solid-body rotation of the fluid. } 
\label{fig:tidal_flow_frame} 
\end{figure}

Let us assume a simple case where those two rotations take place in the same plane, the corresponding situation being depicted in figure \ref{fig:tidal_flow_scheme}.
In such a configuration, the flow inside the core is not a solid body rotation. 
We aim at determining the inviscid flow created by the differential rotation of the planet and its moon; it must satisfy the Euler equation and the non-penetration boundary conditions at the edge of the ellipsoidal container. 
We use the method introduced by \cite{hough_xii._1895}. 
First we note that the derivation is facilitated when carried out in the frame of reference in which the boundary shape does not change over time. 
Then, we look for a uniform vorticity solution which directly satisfies the boundary conditions. 
This is achieved by first using a rescaled system of coordinates $\tilde{\mbf{X}} = (\tilde{X},\tilde{Y},\tilde{Z})$ which transforms the ellipsoid into a sphere, the velocity $\mbf{U} = (U,V,W)$ being also transformed accordingly:
\begin{equation}
\left\lbrace
\begin{array}{rl}
\tilde{X} &= X/a \\
\tilde{Y} &= Y/b \\
\tilde{Z} &= Z/c  
\end{array}\right.
 ~~\mbox{and} ~~~
 \left\lbrace
\begin{array}{rl}
\tilde{U} &= U/a \\
\tilde{V} &= V/b \\
\tilde{W} &= W/c  
\end{array}\right.
\end{equation}
and then looking for a vector $\mbf{\omega} (t)$ such that:
\begin{equation}
\label{eq:ansatz}
\tilde{\mbf{U}} = \mbf{\omega} (t) \times \tilde{\mbf{X}}.
\end{equation} 
With this ansatz, the transformed velocity field is at each time $t$ a solid-body rotation which necessarily satisfies the non-penetration boundary conditions in the sphere. 
As a consequence, the flow $\mbf{U}$ transformed back into the original coordinates also satisfies the boundary conditions. 
Such a flow is divergence-free and has a uniform vorticity $\mbf{\varpi}$ which writes: 
\begin{equation}
\mbf{\varpi} = \mbf{\nabla} \times \mbf{U}=
\begin{bmatrix}
\displaystyle
\left(\frac{c}{b} + \frac{b}{c} \right) \omega_x \\[1em]
\displaystyle
\left(\frac{c}{a} + \frac{a}{c} \right) \omega_y \\[1em]
\displaystyle
\left(\frac{a}{b} + \frac{b}{a} \right) \omega_z 
\end{bmatrix} ~.
\end{equation} 
We look for a steady solution in the frame orbiting with the Moon at a rate $n \mbf{e}_z$ ---see figure \ref{fig:tidal_flow_scheme}.
Taking into account the Coriolis acceleration due to the rotation of the frame of reference, the stationary vorticity equation derived from the Euler equation reads: 
\begin{equation}
\label{eq:vorticity_equation}
(\mbf{U} \cdot \mbf \nabla) \mbf{\varpi}  = \left( (\mbf{\varpi} + 2 n \mbf{e}_z )\cdot \mbf{\nabla} \right)
\mbf{U} ~.
\end{equation}
%
%
As $\mbf{\varpi}$ is space-independent, the left hand side of this 
equation vanishes. 
The equations on the components of the vorticity are therefore:
\begin{equation}
\label{eq:vorticity}
\left\lbrace
\begin{array}{rl}
\displaystyle\left[ \frac{1}{c} \left(\frac{a}{b} + \frac{b}{a} \right) -  \frac{1}{b} \left(\frac{c}{a} + \frac{a}{c} \right)\right]  \omega_y \omega_z + 2 \frac{1}{c} n \omega_y &= 0 \\[1em]
\displaystyle
\left[ \frac{1}{a} \left(\frac{c}{b} + \frac{b}{c} \right) -  \frac{1}{c} \left(\frac{a}{b} + \frac{b}{a} \right)\right]  \omega_x \omega_z - 2 \frac{1}{c} n \omega_x &= 0 \\[1em]
\displaystyle
\left[ \frac{1}{b} \left(\frac{c}{a} + \frac{a}{c} \right) -  \frac{1}{a} \left(\frac{c}{b} + \frac{b}{c} \right)\right] \omega_x \omega_y &= 0  ~.
\end{array} 
\right.
\end{equation}
The last equation prescribes either $\omega_x$ or $\omega_y$ to be equal to $0$. Let us assume $\omega_y = 0$ (a similar reasoning is possible for $\omega_x = 0$): the first equation is then directly satisfied. The second equation either prescribes a $\omega_z$ proportional to $n$ for a non-zero $\omega_x$, or a zero $\omega_x$. However, the former possibility does not account for the global rotation of the planet at rate $\Omega$, which must be part of the solution. 
The physical solution is therefore obtained for $\omega_x = 0$.\footnote{This condition is in any case the only one when we consider the viscous problem where the flow solution of (\ref{eq:vorticity}) must reconnect to the rotating solid mantle through a thin boundary layer, see for instance \cite{tilgner_8.07_2007}.} 
%
To constrain the value of $\omega_z$, we consider that the vorticity of the fluid must match the planetary vorticity, \ie~: 
\begin{equation}
\label{eq:vorticity_match}
\left(\frac{a}{b} + \frac{b}{a} \right) \omega_z  = 2 (\Omega - n)~~ \Longleftrightarrow \omega_z = \frac{2ab}{a^2 + b^2}  (\Omega - n)
\end{equation}
which finally gives the following steady flow driven by tides: 
\begin{equation}
\begin{bmatrix}
U \\
V \\
W 
\end{bmatrix} 
= 
\frac{2ab}{a^2 + b^2} ~ (\Omega - n)~
\begin{bmatrix}
\displaystyle
-~\frac{a}{b}~ Y \\[1em]
\displaystyle
\frac{b}{a}~ X  \\[1em]
0 
\end{bmatrix}
\end{equation}
Lastly, we can introduce the ellipticity of the deformation $\beta = (a^2 - b^2)/(a^2 + b^2)$; the base flow then writes into a simpler and more compact form: 
\begin{equation}
\label{eq:tidal_base_flow_n}
\mbf{U} =  (\Omega - n)
\begin{bmatrix}
 0 & -1 - \beta & 0 \\
 1 - \beta & 0 & 0 \\
 0 & 0 & 0 
\end{bmatrix}
\begin{bmatrix}
X \\
Y \\
Z 
\end{bmatrix}~.
\end{equation}
Note that this last form is quite meaningful as it corresponds to the superposition of a circular vortex and a strain, which is a configuration known to be unstable \citep{pierrehumbert_universal_1986,bayly_three-dimensional_1986,waleffe_threedimensional_1990}.
Moreover, in the frame rotating with the planet, the flow $\mbf{U}$, denoted $\mbf{U}^\Omega$, is:
\begin{equation}
\label{eq:tidal_base_flow_O}
\mbf{U}^\Omega =  - \beta \gamma
\begin{bmatrix}
 \sin (2 \gamma t)  & \cos(2 \gamma t) & 0 \\
 \cos(2 \gamma t)  & -  \sin (2 \gamma t) & 0 \\
 0 & 0 & 0 
\end{bmatrix}
\begin{bmatrix}
x \\
y \\
z  
\end{bmatrix}~~~~~ \mbox{with:} ~~ \gamma = \Omega-n 
\end{equation}
which can be retrieved from (\ref{eq:tidal_base_flow_n}) via a rotation of coordinates and a velocity composition. 
The two fields (\ref{eq:tidal_base_flow_n}) and (\ref{eq:tidal_base_flow_O}) are shown in figure \ref{fig:tidal_flow_frame}.
This last way of writing the tidal flow is even more meaningful as it highlights the time periodicity of the tidal excitation: the tidal frequency is twice the differential rotation between the planet and the moon, which reflects the symmetry of the tidal bulge respective to the plane $(OYZ)$.  
For $n\ll \Omega$, this is tantamount to undergoing two tidal rises a day as on the Earth where $n/\Omega \sim 1/27$.
Besides, the tidal flow amplitude is proportional to the ellipticity of the deformation.
Tidal excitation is a small perturbation to the planetary solid-body rotation, of relative amplitude $10^{-7}$ on Earth for instance.
Nevertheless, we show later in this chapter that this repetitive excitation, although of small amplitude, is able to excite turbulent flows.

\subsection{Perturbation of the rotation rate: libration}

In the previous section, we have introduced the tidal potential and its consequences on the shape of planetary cores. 
Tides not only induce distortion, they also alter the rotation of planets. 
We review in the next two sections typical perturbations of the rotation rate and rotation axis, and combine them with the tidal distortion to derive the corresponding core flows. 

\subsubsection{Libration of moons}
\begin{figure}
\centering
\includegraphics[width=\linewidth]{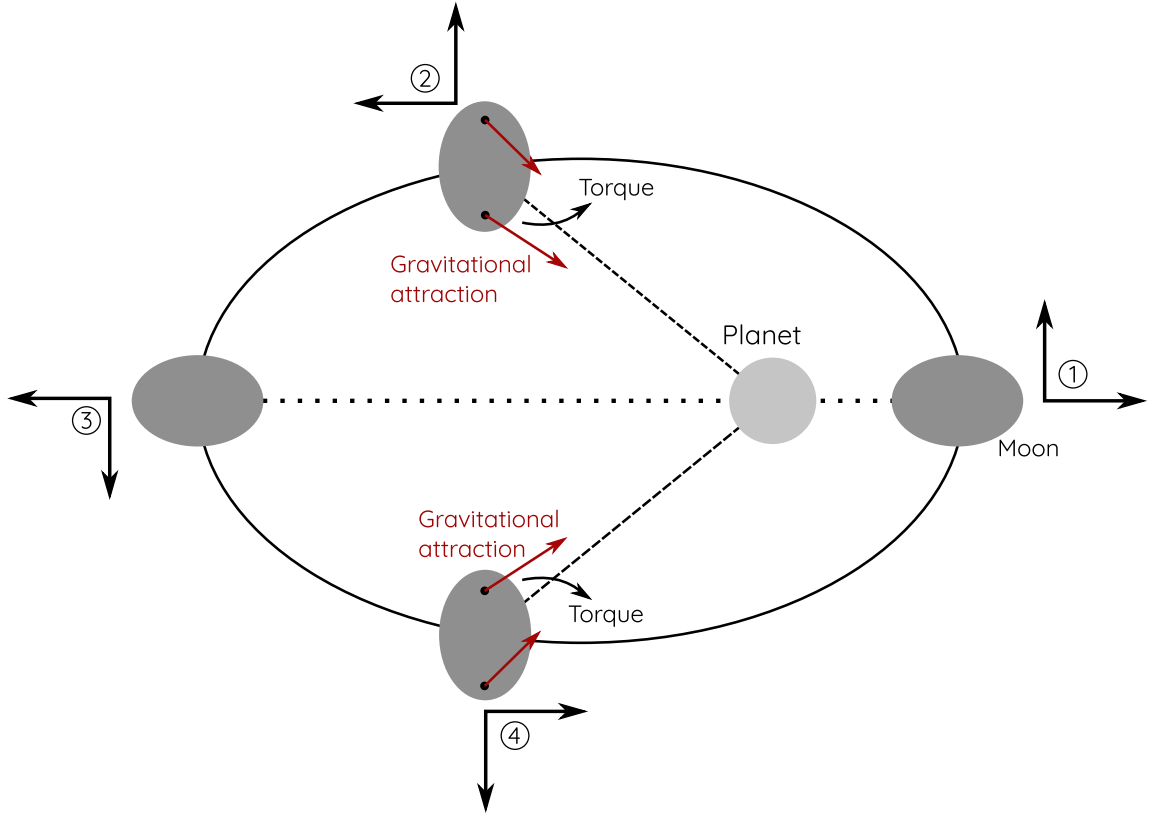}
\caption{Schematic diagram of the excitation of libration on a moon with an eccentric orbit in spin-orbit resonance. Its bulge is frozen and follows the rotation of the moon instead of staying aligned towards the parent planet. The four snapshots separate the orbit and the rotation in four equal periods.}
\label{fig:tial_libration}
\end{figure}

Physical longitudinal libration, hereafter called libration, is the oscillation of the rotation rate of an astrophysical body without change in its rotation axis. 
This kind of motion is excited by tidal interaction between the considered body's tidal bulge and its parent planet or star.
One common situation where libration is observed is presented in figure \ref{fig:tial_libration}: 
a moon is synchronized in a spin-orbit resonance along an elliptical orbit, meaning that its orbital rate matches its mean rotation rate (like our Moon, which always shows us the same side). 
This is due to tidal dissipation inside the rocky mantle of such bodies which despins them from any initial rotation rate into this particular equilibrium \citep{rambaux_tides_2013}.
Moreover, because of the rigidity of the moon, its tidal bulge is in general not exactly aligned with the parent body. 
The figure \ref{fig:tial_libration} presents the extreme situation where the bulge is frozen and follows the rotation of the moon instead of staying aligned with the planet.
This happens for instance in the case of the Earth's Moon which has a large fossil bulge which has not relaxed and is not induced by tidal interaction anymore but still persists. 
As depicted in figure \ref{fig:tial_libration}, this misalignment and the difference of gravitational attraction between the two sides of the bulge create a torque which tends to accelerate the rotation rate at \circled{2} and to decelerate the rotation rate at \circled{4}. 
The rotation rate is therefore perturbed around a mean with the same period as the orbit. 
The situation shown in figure \ref{fig:tial_libration} is not the only one leading to libration oscillations. 
In planetary systems with many satellites such as as the Jovian and Saturnian systems, a body not only interacts with its parent planet but also with all the other moons.
Some of them are in what is called a Laplace resonance, which is a stable situation where the orbit rates of several moons are multiples of 
one another \citep{rambaux_tides_2013}.

\subsubsection{Libration-driven flows}
We seek now to determine the flow driven by libration and assume for simplicity's sake a purely rigid tidal bulge. 
The total rotation rate including libration can be written as:
\begin{equation}
\label{eq:libration}
\mbf{\Omega} = \Omega_0 \left( 1 + \varepsilon \sin(\omega_\ell t) \right)  \mbf{e}_z
\end{equation}
where $\varepsilon$ is the relative variation of the rotation rate and $\omega_\ell$ is the libration frequency.
Observed from the frame rotating at rate $\Omega_0$, the moon and its bulge oscillate with a frequency $\omega_\ell$ and with an amplitude angle $\varepsilon \Omega_0 / \omega_{\ell}$.
Although in the example of spin-orbit synchronization presented in figure \ref{fig:tial_libration} $\omega_{\ell} = \Omega_0$, the libration frequency can take any value due to tidal interactions with many bodies as it the case in the Jovian and Saturnian systems. 
To determine the libration-driven flow, we use the same method as for the tidal base flow: we look for a flow $\tilde{\mbf{U}} = \mbf{\omega} (t) \times \tilde{\mbf{X}}$ --- see definition (\ref{eq:ansatz})--- in the frame where the boundary stands still, \ie~the librating frame. 
We use the same notation as before: in the frame $(OXYZ)$ the boundary is still ---therefore it is the librating frame--- and $(Oxyz)$ is the mean rotation frame.  
In the $(OXYZ)$ frame the equation for the vorticity $\mbf{\varpi}$ writes:
\begin{equation}
\label{eq:vorticity_libration}
\p_t \mbf{\varpi}  + (\bu \cdot \mbf \nabla) \mbf{\varpi}  = \left( (\mbf{\varpi} + 2 \mbf{\Omega})\cdot \mbf{\nabla} \right) \mbf{u} + 2 \dot{ \mbf{\Omega}}
\end{equation}
where we include the time dependence of the flow ---which oscillates at the libration frequency--- and a last term corresponding to the Poincar\'e's acceleration. 
In its expanded form, the equation (\ref{eq:vorticity_libration}) yields:
\begin{equation}
\label{eq:vorticity_libration_expanded}
\left\lbrace
\begin{array}{rl}
\displaystyle \left(\frac{c}{b} + \frac{b}{c} \right) \dot{\omega}_x ~&=~ \displaystyle
 \left[ \frac{a}{c} \left(\frac{a}{b} + \frac{b}{a} \right) -  \frac{a}{b} \left(\frac{c}{a} + \frac{a}{c} \right)\right]  \omega_y \omega_z + 2 \frac{a}{c} \Omega \omega_y \\[1em]
\displaystyle
\left(\frac{c}{a} + \frac{a}{c} \right) \dot{\omega}_y ~&=~
\displaystyle
\left[ \frac{b}{a} \left(\frac{c}{b} + \frac{b}{c} \right) -  \frac{b}{c} \left(\frac{a}{b} + \frac{b}{a} \right)\right]  \omega_x \omega_z - 2 \frac{b}{c} \Omega \omega_x \\[1em]
\displaystyle
\left(\frac{a}{b} + \frac{b}{a} \right) \dot{\omega}_z ~&=~
\displaystyle
\left[ \frac{c}{b} \left(\frac{c}{a} + \frac{a}{c} \right) -  \frac{c}{a} \left(\frac{c}{b} + \frac{b}{c} \right)\right] \omega_x \omega_y + 2 \dot{\Omega}  ~.
\end{array} 
\right.
\end{equation}
Assuming $\omega_x$ and $\omega_y$ are zero, the last equation relates the temporal variation of $\omega_z$ to the Poincar\'e acceleration:
\begin{equation}
\left(\frac{a}{b} + \frac{b}{a} \right) \dot{\omega}_z ~=~ 2 \dot{\Omega} ~~ \Longleftrightarrow ~~ \omega_z = \frac{2ab}{a^2 + b^2} \Omega_0 \varepsilon \sin(\omega_\ell t) + \overline{\mathrm{cst}}~.
\end{equation}
Considering a synchronized body in the librating frame, the constant component of the vorticity must be equal to zero. 
The resulting base flow can be written in the librating frame in terms of axes lengths or the ellipticity $\beta$ defined earlier: 
\begin{align*}
\label{eq:libration_base_flow_librating_frame}
\mbf{U} &= \frac{2ab}{a^2 + b^2} \Omega_0 \varepsilon \sin(\omega_\ell t)
\begin{bmatrix}
 0 & -a/b & 0 \\
 b/a & 0 & 0 \\
 0 & 0 & 0 
\end{bmatrix}
\begin{bmatrix}
X \\
Y \\
Z 
\end{bmatrix}~ 
\\&= \Omega_0 \varepsilon \sin(\omega_\ell t)
\begin{bmatrix}
 0 & -1 - \beta & 0 \\
 1-\beta & 0 & 0 \\
 0 & 0 & 0 
\end{bmatrix}
\begin{bmatrix}
X \\
Y \\
Z 
\end{bmatrix}~ 
\end{align*}
The libration flow written in the mean rotation frame ---which requires doing velocity composition and coordinates change--- writes, at the lowest order in $\varepsilon$:
\begin{equation}
\label{eq:libration_base_flow}
\mbf{U}^{\Omega} ~=~ \Omega_0 \varepsilon \beta \sin(\omega_\ell t)
\begin{bmatrix}
 0 & 1  & 0 \\
 1 & 0 & 0 \\
 0 & 0 & 0 
\end{bmatrix}
\begin{bmatrix}
x \\
y \\
z 
\end{bmatrix}~. 
\end{equation}
Note that at $t = \pi/(2\omega_\ell)$ the structure of the flow is exactly similar to (\ref{eq:tidal_base_flow_n}) and (\ref{eq:tidal_base_flow_O}): snapshots of the velocity in the libration case are the sames as figure \ref{fig:tidal_flow_frame} left for the librating frame and figure \ref{fig:tidal_flow_frame} right for the mean rotation frame. 
Generally, seen from the frame where the bulge is stationary, a fluid particles rotates around elliptical streamlines in the tidal case and oscillates along elliptical streamlines in the libration case. Seen from the mean rotation frame, the strain field in the tidal case rotates with a constant amplitude, whereas the strain field in the libration case has a stationary spatial structure but an oscillating amplitude.

\subsection{Perturbations of the rotation axis: precession}

\subsubsection{Precession and equatorial bulge}

\begin{figure}
\centering
\includegraphics[width=\linewidth]{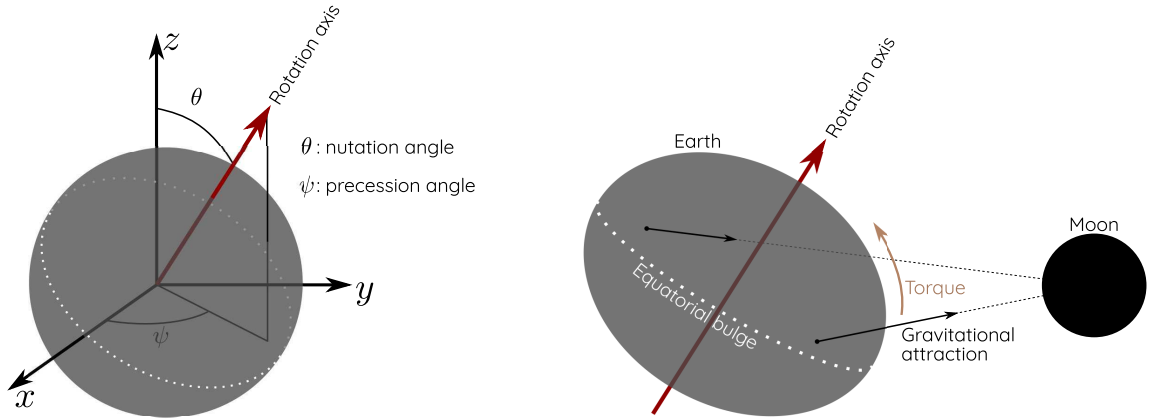}
\caption{\textbf{Left}: illustration of the two motions of the rotation axis which can be decomposed in nutation ---variations of $\theta$--- and precession ---variations of $\psi$. $(xOy)$ is the ecliptic plane.
\textbf{Right}: schematic diagram of the cause of the Earth's precession, which can be applied to other bodies. The gravitational force of a companion body ---here the Earth's moon--- creates a torque which tends to align the equatorial bulge created by rotation with the Moon. The gyroscopic effect due to  rapid spinning converts this torque into a precession of the rotation axis. }
\label{fig:precession_nutation}
\end{figure}

The spin of planets around their axes is in general predominant compared to any other motion in the barycentric frame of reference. 
Gravitational interactions can force the spinning axis to change over time, this motion been decomposed into precession and nutation ---see figure \ref{fig:precession_nutation}.a. 
We rather dwell hereafter on precession, which is a direct consequence of the existence of an equatorial bulge induced by the planet spin and the associated centrifugal force.
As depicted in figure \ref{fig:precession_nutation}.b, a neighboring body exerts a net gravitational torque which tends to align the equatorial bulge with the body. 
A gyroscopic effect then applies and converts this torque into precession, \ie~constant increase of the angle $\psi$ with constant $\theta$. 
This is the case for the Earth which precesses over a period of $26000$ years with an angle $\theta \sim 23.5^\circ$ because of gravitational interactions with the Moon and the Sun.  

\subsubsection{Precession-driven flow}

\begin{figure}
\centering
\includegraphics[width = 0.9\linewidth]{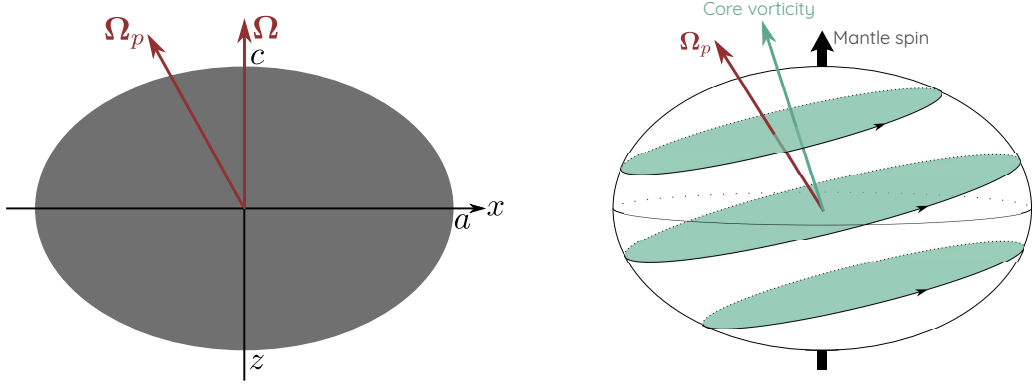}
\caption{\textbf{Left}: schematic cartoon of a precessing spheroid and the different rotation vectors as seen from the precession frame of reference. The fast rotation vector $\mbf{\Omega}$ of the mantle surrounding the  liquid core is along the $z$ axis. The frame rotates at rate $\Omega_p \mbf{e}_z$ so that the precession and the rotation vectors remain in the same plane.\textbf{Right}: Visualization of the precession driven flow whose vorticity is along an intermediate direction between the precession vector and the mantle fast spinning.}
\label{fig:precession_scheme}
\end{figure}

As for libration and tides, we would like to compute the base flow driven in a precessing planetary core.
To fully expand the vorticity equation, we assume for simplicity's sake that the core is spheroidal, \ie~that $a = b$.
As seen earlier, tides generally impose $a \neq b$: this case is also tractable, yet more complicated \citep{cebron2010tilt}. 
With this symmetry, we carry out the computation of the precession-driven flow in the precessing frame, which is the frame in which the rotation axis of the solid mantle is constant equal to $\Omega \bold{e}_z$ ---see figure \ref{fig:precession_scheme}.
This frame rotates at the precession rate $\Omega_p$ around the $z$ axis so that both the rotation vector $\mbf{\Omega}$ and the precession vector $\mbf{\Omega}_p$  remain in the same plane $(Oxz)$, as shown in figure \ref{fig:precession_scheme}.
Note also that because we assume $a=b$, the topology of the core outer boundary remains unchanged although it rapidly rotates around the $z$ axis.

%
We then use exactly the same method as previously and the equations to solve to determine the vector $\mbf{\omega}$ are:
\begin{equation}
\label{eq:vorticity_libration_expanded}
\left\lbrace
\begin{array}{rl}
\displaystyle
\left(\frac{c}{a} + \frac{a}{c} \right) \dot{\omega}_x ~&=~
\displaystyle
\left[ \frac{a}{c} -  \frac{c}{a} \right]  \omega_y \omega_z + 2 \frac{a}{c} \Omega_{pz} \omega_y \\[1em]
\displaystyle
\left(\frac{c}{a} + \frac{a}{c} \right) \dot{\omega}_y ~&=~
\displaystyle
\left[  \frac{c}{a}  -  \frac{a}{c} \right]  \omega_x \omega_z - 2 \frac{a}{c} \Omega_{pz} \omega_x + 2 \Omega_{px} \omega_z  \\[1em]
2 \dot{\omega}_z ~&=~
\displaystyle
- 2 \frac{c}{a} \Omega_{px} \omega_y.
\end{array} 
\right.
\end{equation}
Note that these equations include a Coriolis force associated to the precession rate. 
We look for a steady solution, and the last equation implies that $\omega_y = 0$, which is also consistent with the first one: the rotation axis of the liquid core is in the same $(Oxz)$ plane as the precession axis (see figure \ref{fig:precession_scheme} right). 
The second equation yields: 
\begin{equation}
\label{eq:poincare_precession}
\frac{\omega_x}{\omega_z} ~=~  \frac{\Omega_{px}}{ \displaystyle\frac{a}{c} \Omega_{pz} + \frac{a^2 - c^2}{2ac} \omega_z}.
\end{equation}
As noted in \cite{tilgner_8.07_2007} and \cite{le_bars_flows_2015}, the inviscid theory does not allow to relate the $z$ component of the core vorticity to the mantle's spin in order to close the problem: this can be achieved only via the introduction of viscous effects. 
The interested reader can refer to the key study of \cite{busse_steady_1968} and discussions in the reviews of \cite{tilgner_8.07_2007} or \cite{le_bars_flows_2015}.
Here, we simply give first order considerations in the limit of small precession rates. In the absence of precession ($\Omega_p =0$), one expect the fluid to rotate with the mantle, hence $\omega_z = \Omega$. By symmetry between prograde and retrograde precession for the vertical vorticity at small precession rate, one can also expect the first correction of $\omega_z$ to appear at order 2 only, \ie~$\omega_z = \Omega + O(\Omega_p^2)$. Then the horizontal vorticity is given at first order by 
\begin{equation}
\label{eq:poincare_sphere}
\frac{\omega_x}{\Omega} ~=~  \frac{\Omega_{px}}{ \displaystyle\frac{a}{c} \Omega_{pz} + \frac{a^2 - c^2}{2ac} \Omega}
\end{equation}
and the base flow written the precessing frame of reference is: 
\begin{equation}
\label{precession_flow}
\mbf{U}_b^p = 
\begin{bmatrix}
-\Omega Y \\[0.5em]
\Omega X \\[0.5em]
 \displaystyle\frac{c}{a} \frac{\Omega_{px}}{ \displaystyle\frac{a}{c} \Omega_{pz} + \frac{a^2 - c^2}{2ac} \Omega} \Omega
\end{bmatrix}~ ~.
\end{equation}
It is represented schematically in figure \ref{fig:precession_scheme} left. 
For instance in the case of the Earth which has a rotation axis titled of $23.5^\circ$ compared to the ecliptic plane, a polar flattening of about $1/400$ and a precession period of about 26000 years, the angle between the mantle's and the core's spinning axes is about $0.001^\circ$. 
We note that equation (\ref{eq:poincare_sphere}) obtained in a very simplified approach predicts a ``resonance'' (i.e. a divergence of $\omega_x$) for specific combinations of shape and precession parameters. A full calculation including viscosity exhibits a range of phenomena including resonance and bistability \citep{cebron2015bistable}.

\section{Instabilities driven by mechanical forcings: from parametric resonance to turbulence}
\vspace*{1cm}

In the preceding section, we have determined the primary response of liquid cores to tidal deformation and rotation perturbations such as libration and precession. 
We show in this section that the repetitive action of these forcings leads to bulk instabilities. 
The core destabilization is due to the parametric resonance of inertial waves with the periodic primary flow. 
This section begins with a presentation of parametric resonance through the example of the Botafumeiro, a famous length-varying pendulum.  
We then introduce inertial waves and the possibility of resonance with tidal perturbations.

\subsection{Parametric sub-harmonic resonance of a pendulum}
\label{sec:pendulum}

\subsubsection{O Botafumeiro}

One of the most striking example of parametric resonance that can be found around the globe is probably \textit{O Botafumeiro} in Santiago de Compostella's Cathedral. 
It is a 54~kg thurible ---a metal censer--- that hangs on the top of the cathedral's dome. 
The length of the rope can be changed over time around its mean length of 21.5~m by a group of holders. 
To spread incense in the cathedral, the holders first let the thurible swing with a small angle. 
Each time the censer goes up, \textit{i.e.} twice per period, they slightly pull down the rope to shorten the swing length, and let it increase again as the swing goes down. 
With this twice per period excitation, they manage to swing the censer with very large amplitude in a short time, up to a height of 20.6~m. The velocity at the lowest point of the oscillation reaches 68~km/h. 
\begin{figure}
\centering
\includegraphics[width=0.28\linewidth]{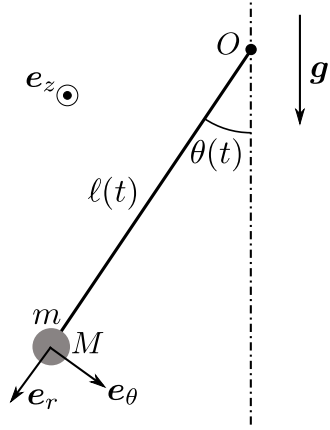}
\caption{Model for \textit{O Botafumeiro}: a length varying pendulum. The function $\ell(t)$ is prescribed to periodically vary around a mean length $\ell_0$ according to (\ref{eq:pendulum_length}).}
\label{fig:length_varying_pendulum}
\end{figure}

\subsubsection{Mechanical study of the length varying pendulum}
As shown in figure \ref{fig:length_varying_pendulum}, we model \textit{O Botafumeiro} by a pendulum whose length $\ell$ varies over time: 
\begin{equation}
\label{eq:pendulum_length}
\ell (t) ~=~ \ell_0 ( 1 + \eta \sin(\omega t) )
\end{equation}
where $\eta$ is a small parameter accounting for the small modulation of the relative length of the swing. 
We study the motion of the point object $M$ of mass $m$ at the end of the rope of length $\ell (t)$. We assume that the referential bound to $O$ is Galilean and that the rope is mass-less.
The velocity $\mbf{v}$ of $M$ writes:
\begin{equation}
\mbf{v} ~=~ \ell \dot{\theta} \mbf{e}_\theta + \dot{\ell} \mbf{e}_r
\end{equation}
and the angular momentum  $L_z$ of $M$ with respect to axis $(O,\mbf{e}_z)$ is:
\begin{equation}
L_z ~=~  m \ell^2 \dot{\theta} ~.
\end{equation}
As gravity applies a torque $- m \ell g \sin \theta$, the conservation of angular momentum yields:
\begin{equation}
\label{eq:obotafumeiro_equation}
\ddot{\theta} + \frac{2 \dot{\ell}}{\ell} \dot{\theta} + \frac{g}{\ell} \sin \theta = 0~.
\end{equation}
Introducing the ansatz (\ref{eq:pendulum_length}) and $\omega_0^2 = g/\ell_0$ finally leads to:
\begin{equation}
\label{eq:parametric_equation}
\ddot{\theta} + \eta \frac{2 \omega \cos(\omega t)}{1 + \eta \sin(\omega t)} \dot{\theta} + \frac{\omega_0^2}{1 + \eta \sin(\omega t)} \sin \theta ~=~0,
\end{equation}
where $\eta$ and $\omega$ are two control parameters, the relative variation of the rope length and the frequency of these variations.

We can already qualitatively predict at which frequency the length must be varied for optimal oscillation by looking at equation (\ref{eq:obotafumeiro_equation}): indeed, comparing to the classical pendulum equation where $\ell$ is constant but still accounting for $\dot{\ell}$, the second term in (\ref{eq:obotafumeiro_equation}) appears in place of a viscous damping of type $\nu \dot{\theta}$. 
Here however the sign of the viscosity depends on the sign of $\dot{\ell}$. 
In particular, shortening the length of the rope ---\ie~$\dot{\ell}<0$--- corresponds to a ``negative viscosity'', which encourages the motion;  $\dot{\ell}>0$ corresponds to a classical positive damping.
For the holders to input maximum energy into the system, the length must thus be shortened when the angular velocity $|\dot{\theta}|$ is maximum, \ie~when the pendulum is at its lowest position, and increased when $\dot{\theta}=0$, \ie~when it is at its highest position. 
Therefore, two antagonistic moves must be operated during half the period of the pendulum: the frequency of the excitation must be the double of the free oscillation. 
This will be formally proved in the following.

\subsubsection{Asymptotic analysis}
\begin{figure}
\centering
\includegraphics[width=\linewidth]{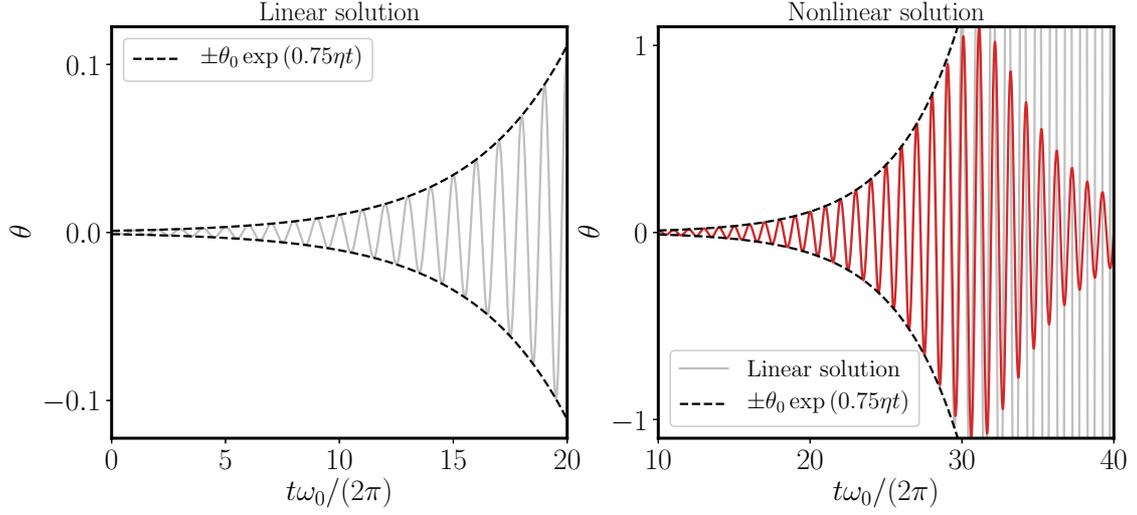}
\caption{\textbf{Left}: linear solution corresponding to the expression (\ref{eq:resonance_total_solution}) for $\eta = 5 \times 10^{-2}$, $\omega_0 = 1$, $\omega = 2$ and $\theta_0 = 1 \times 10^{-3}$.  \textbf{Right:} fully non-linear solution of equation (\ref{eq:parametric_equation})  (in red) and comparison with the linear solution (in gray).  }
\label{fig:parametic_resonance}
\end{figure}

To analytically study the length-varying pendulum described by (\ref{eq:parametric_equation}) and get the main physical properties of this system, we carry out an asymptotic study in the limit of small length variations $\eta \ll 1$ and small angle $\theta \ll 1$. 
The equation (\ref{eq:parametric_equation}) can then be expanded into:
\begin{equation}
\label{eq:parametric_first_order}
\ddot{\theta} + \omega_0^2 \theta = \eta \left[ -2 \omega \cos(\omega t) \dot{\theta} + \omega_0^2 \sin(\omega t) \theta \right]
\end{equation}
where the length variation appears at order one as a forcing. 
The system has two typical time scales. 
The first one is $1/\omega_0$, the natural or free frequency of pendulum oscillation. This is the fast timescale of the system. We further assume here that the forcing period $1/\omega$ is of the same order of magnitude as $1/\omega_0$, because as seen just before, this is where interesting physics is expected. 
Then because of the forcing, the system also evolves over a slow timescale $1/(\eta \omega_0)$. 
We thus look for a solution to (\ref{eq:parametric_first_order}) where these two timescales are included and decoupled, \ie~a function $\theta$ which depends on $\tau = t$ and $T = \eta t$ such that:
\begin{equation}
\label{eq:twotiming_ansatz}
\theta (\tau,T) = \left(f_0(\tau)+ \eta f_1(\tau) \right) F(T) ~. 
\end{equation}
In this two-timing framework, the total time derivatives are expanded as partial derivatives according to: 
\begin{equation}
\label{eq:twotiming_partial}
\left\lbrace
\begin{array}{rl}
\ddroit{}{t} &= \displaystyle \frac{\partial}{\partial \tau} + \eta \frac{\partial}{\partial T} \\[1em]
\ddroit{^2}{t^2} &= \displaystyle\frac{\partial^2}{\partial \tau^2} + 2 \eta   \frac{\partial^2}{\partial T \partial \tau}
\end{array}
\right.
\end{equation}
where we have kept only the terms up to order 1. 
Taking into account the ansatz (\ref{eq:twotiming_ansatz}) and (\ref{eq:twotiming_partial}) yields to an order zero and an order one equations such that:
\begin{equation}
\label{eq:twotiming_pendulum}
\left\lbrace
\begin{array}{rl}
\ddroit{^2 f_0}{\tau^2} + \omega_0^2 f_0 &= 0 \\[1em]
\ddroit{^2 f_1}{\tau^2} + \omega_0^2 f_1 &= -2 \omega \cos(\omega \tau) f_0' + \omega_0^2  f_0 \sin(\omega \tau) - 2    \displaystyle \frac{ f_0' F'}{F} ~.
\end{array}
\right.
\end{equation}
The solution to the first equation is straightforward, and using complex solutions: 
\begin{equation}
\label{eq:f0_solution}
f_0 (\tau) ~=~ A e^{i \omega_0  \tau} + B e^{-i \omega_0 \tau}~.
\end{equation}
We then input this solution in the right hand side (RHS) of the second equation in (\ref{eq:twotiming_pendulum}) and expand it to find the following Fourier decomposition: 
\begin{equation}
\label{eq:twotiming_RHS}
\begin{array}{rl}
\ddroit{^2 f_1}{\tau^2} + \omega_0^2 f_1 ~=~  &e^{i (\omega_0 + \omega )\tau} \left[ -A i \omega_0 \omega + A \displaystyle\frac{\omega_0^2}{2i} \right] \\[1em]
 +  
&e^{i ( \omega_0-\omega  )\tau} \left[- A i \omega_0 \omega - A \displaystyle\frac{\omega_0^2}{2i} \right] \\[1em]
+
&e^{i (-\omega_0 + \omega )\tau} \left[ B i \omega_0 \omega + B\displaystyle \frac{\omega_0^2}{2i} \right] \\[1em]
 +
&e^{-i (\omega_0 + \omega )\tau} \left[ B i \omega_0 \omega - B \displaystyle\frac{\omega_0^2}{2i} \right] \\[1em]
 +
&e^{i \omega_0 \tau} \left[ -2 i A  \displaystyle\frac{F'}{F} \omega_0 \right]
+  e^{-i \omega_0 \tau} \left[ 2 i B  \displaystyle\frac{F'}{F} \omega_0 \right] ~. \\[1em]
\end{array}
\end{equation}
The frequency of the two terms appearing in the last line is the same as the eigen frequency of the harmonic oscillator in the left hand side (LHS) of (\ref{eq:twotiming_RHS}) and should give rise to divergence of $f_1$. 
Those contributions to the LHS are called ``secular terms'' as they excite a long term growth of the solution. 
However, the Taylor expansion in the ansatz (\ref{eq:twotiming_ansatz}) requires $f_1$ to remain bounded over time for the calculation to remain valid: secular terms must therefore be canceled. 
In general, a first possibility is to impose $F' = 0$, but this cannot explain the amplitude increase observed in the case of \textit{O Botafumeiro}. 
A more interesting solution arises when the excitation frequency $\omega$ is adequately chosen. 
When $$\omega = 2 \omega_0$$ the second and third lines then also have a frequency of $\pm \omega_0$ and give a more complex condition for $f_1$ to remain bounded. 
With this particular condition, canceling the secular terms leads to
\begin{equation}
\left\lbrace
\begin{array}{rl}
 -\displaystyle\frac{3}{2} \omega_0 A + 2 \frac{F'}{F} B ~&=~ 0 \\[1em]
 -\displaystyle 2 \frac{F'}{F} A + \frac{3}{2} \omega_0 B ~&=~ 0 ~. 
\end{array}
\right.
\end{equation}
In order to avoid the simple solution $A=B=0$ which does not model the amplitude growth, the determinant of the above system must be zero. 
Canceling the determinant imposes:  
\begin{equation}
\label{eq:resonance_conclusion} 
F(T) \propto  e^{\pm \frac{3}{4} \omega_0  T}.
\end{equation}
Under these conditions, it is straightforward that for the exponentially growing branch $A = B$.
The total solution at lowest order, in terms of time $t$ and only considering the growing solution for a pendulum released at $t=0$ from angle $ \theta_0$ with no initial velocity, is therefore: 
\begin{equation}
\label{eq:resonance_total_solution}
\theta (t) ~=~ \theta_0~ \cos(\omega_0 t)~ \exp\left(\displaystyle\frac{3}{4} \eta \omega_0 t \right) ~.
\end{equation}
Such a resonance process is called ``parametric sub-harmonic resonance" as it happens when the excitation frequency is twice the free oscillation frequency. 
From a tiny perturbation, provided it is repetitive and has the adequate frequency, it leads to a drastic increase of the oscillations' amplitude. 
This solution is represented in figure \ref{fig:parametic_resonance} left and superimposed to the fully nonlinear solution obtained via numerical resolution of equation (\ref{eq:parametric_equation}) in figure \ref{fig:parametic_resonance} right. 
The two solutions are in very good agreement at early time, but our linear approach does not capture the collapse of the amplitude observed in the numerical solution. 
As we show in the following paragraph, this is due to nonlinear effects which are not accounted for in our theoretical approach. 

\subsubsection{Saturation of the resonance}
As noted in figure \ref{fig:parametic_resonance} right, the exponential growth computed in the preceding paragraph must come to an end. 
Here, the non-linearity --- the $\sin \theta$ term in (\ref{eq:obotafumeiro_equation})--- causes a collapse of the oscillations.
Indeed, the period of free pendulum oscillations then depends on the amplitude: it increases as the amplitude increases. 
At second order in $\theta$, the relation between period $P$ and amplitude $\theta_m$ is given by Borda's formula \citep{guery-odelin_classical_2010}
\begin{equation}
\label{eq:borda}
P(\theta_m) = \frac{2 \pi}{\omega_0} \left(1 + \frac{\theta_m^2}{16} \right).
\end{equation}
At early times, when the amplitude of the oscillations remains small, the excitation frequency $\omega$ matches the resonance condition $\omega = 2 \omega_0$. 
As the amplitude increases, the frequency of the pendulum decreases and the oscillator is detuned from the excitation. 
This reverses the energy transfer from the parametric excitation to the pendulum. 
As the excitation and oscillation are out of phase, energy is pumped back from the pendulum to the forcing ---the negative viscosity effect becomes a positive viscosity.
The consequence is the global decrease of the amplitude. 
Note lastly that in the case of \textit{O Botafumeiro}, the holders can tune the excitation to the actual period of the thurible, hence maintain large amplitude oscillations over a long time. 

\subsection{Oscillators in planetary cores: inertial waves}
In the following, we draw an analogy between the length-varying pendulum and tidally-driven mechanical forcings in planetary cores. 
We identify what can be regarded as a pendulum inside planetary cores and what acts as a slight and repetitive perturbation.

The oscillating eigenmodes of rotating fluids are the so-called inertial waves. 
They are caused by the restoring action of the Coriolis force. 
This section is a short reminder of how to derive their governing equation and to infer their dispersion relation.
Let us derive a wave equation from the governing equations of incompressible rotating fluids. 
We consider the linear limit of the Euler equation for the velocity $\mbf{u}$ in the rotating frame of reference:
\begin{equation}
\partial_t \bu + 2 \mbf{\Omega} \times \bu = - \mbf{\nabla} p
\end{equation}
The equation governing the vorticity $\mbf{\varpi} = \boldsymbol{\nabla} \times \mbf{u}$ reads:
\begin{equation}
\label{eq:vorticity_linear}
\p_t \mbf{\varpi}   =  2 \left(  \mbf{\Omega}\cdot \mbf{\nabla} \right) \mbf{u} 
\end{equation}
We assume $\mbf{\Omega} = \Omega \mbf{e}_z $ and take the curl the vorticity equation (\ref{eq:vorticity_linear}) to obtain an equation on the velocity only: 
\begin{equation}
\label{eq:pre_poincare_1}
\p_t \left(\mbf{\nabla} \times (\mbf{\nabla} \times \mbf{u} )\right) ~=~ 2 \Omega \p_z \mbf{\varpi}~.
\end{equation}
We differentiate this last equation over time and then substitute (\ref{eq:vorticity_linear}) into (\ref{eq:pre_poincare_1}) which gives:
\begin{equation}
\label{eq:pre_poincare_2}
\partial_{tt} \left(\mbf{\nabla} \times (\mbf{\nabla} \times \mbf{u} )\right) = 4 \Omega^2 \p_{zz} \mbf{u} ~.
\end{equation}
Considering incompressibility,  we finally retrieve the Poincar\'e equation of rotation flows:
\begin{equation}
\label{eq:Poincare_equation}
\partial_{tt} \mbf{\nabla}^2 \mbf{u} + 4 \Omega^2 \p_{zz} \mbf{u} ~=~0~.
\end{equation}
Considering the divergence of the Navier-Stokes equation, one can easily find a similar equation on the pressure field $p$.

\begin{figure}
\centering
\includegraphics[width=0.8\linewidth]{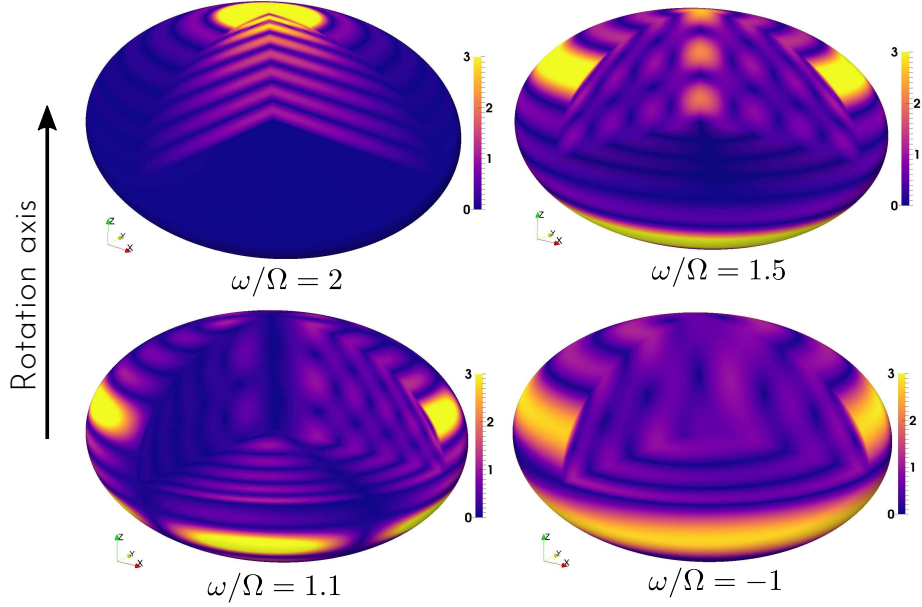}
\caption{Example of four inviscid inertial modes velocity amplitude computed in an ellipsoid with axes $a = 1$, $b = 0.86$ and $c=0.57$ including viscous boundary layers, performed by \cite{vidal_inviscid_2017-1}. It shows equatorial and meridional cuts and the amplitude map at the surface of the ellipsoid. This figure is adapted from \cite{vidal_inviscid_2017-1} figure 6.}
\label{fig:vidal2017} 
\end{figure}

In an hypothetical infinite medium which satisfies translational invariance, the Poincar\'e equation admits plane waves solutions. 
Assuming that $\mbf{u}$ takes the form of a plane wave of vector $\mbf{k}$ and frequency $\omega$ :
\begin{equation}
\label{eq:planar_wave}
\mbf{u}(\mbf{r},t) ~=~ \mbf{u}_0 e^{i(\mbf{k} \cdot \mbf{r} - \omega t)},
\end{equation}
$\mbf{r}$ being the position, the dispersion relation of inertial waves writes:
\begin{equation}
\label{eq:inertial_waves}
\omega^2 = 4 \Omega^2 \frac{k_z^2}{\mbf{k}^2 } ~~~\mbox{\ie} ~~~ \omega = \pm 2 \Omega \cos \xi
\end{equation}
where $\xi$ is the angle between the rotation axis and the wave vector. 
Note that this dispersion relation is peculiar in the sense that wave frequency is not related to the wavelength but to the direction of the wavevector only.
Moreover, the frequency of the waves is bounded between $-2\Omega$ and $2 \Omega$.
In the case of bounded geometry such as spheres or ellipsoids, the translational invariance is lost, but oscillatory solutions to the Poincar\'e equation can still be found.
They are called inertial modes and their derivation for the special case of the sphere can be found in \cite{greenspan_theory_1968}.
Their frequencies remain bounded between $-2 \Omega $ and $2 \Omega$ \citep{greenspan_theory_1968}.
An example of inertial modes computed in an ellipsoidal container is given in figure \ref{fig:vidal2017}.
In the case of periodic mechanical forcing of planets, these inertial modes play the role of the excited oscillators. 

\subsection{Parametric excitation: the case of tidally-driven instabilities}

In section \ref{tidal_flow}, we derived the primary response of a fluid planetary interior to tidal distortion --- see for instance the expressions (\ref{eq:tidal_base_flow_n}) and (\ref{eq:tidal_base_flow_O}); nothing has been said yet regarding the stability over time of this flow.
This section shows that a tidal flow can excite a parametric resonance of inertial waves. 
We give the conditions under which a resonance can happen, and we later briefly present a few ideas on how to quantify its grow rate.  

We investigate the time evolution of perturbations to the tidal base flow, an instability being characterized by an exponential growth of these perturbations. 
In the frame rotating with the planet ---see paragraph \ref{tidal_flow}---, we write the total flow as :
\begin{equation}
\label{eq:ansatz_perturb}
\mbf{U} = \mbf{U}^\Omega + \mbf{u}
\end{equation}
where $\mbf{U}^\Omega$ is the tidal base flow  (\ref{eq:tidal_base_flow_O}) and $\mbf{u}$ is a perturbation ---which is not necessarily small. 
As $\mbf{U}^\Omega$ is a non-linear, viscous solution to the flow in the bulk of the fluid, the Navier-Stokes equations reduce to the non-dimensional form:
\begin{align}
\label{eq:NS_perturb_tidal}
& \partial_t \bu + \mbf{U}^\Omega \cdot  \mbf{\nabla} \bu + \bu \cdot  \mbf{\nabla} \mbf{U}^\Omega + \bu \cdot \mbf{\nabla} \bu  + 2 \mbf{e}_z \times \bu = - \mbf{\nabla} p + E \mbf{\nabla}^2 \bu \\
& \mbf{\nabla} \cdot \bu = 0
\end{align}
where lengths are scaled by the mean core radius $R_c= (a+b)/2$, time by $1/\Omega$, and where we have introduced the Ekman number $E= \nu/(R_c^2 \Omega)$ that compares the effects of viscous and Coriolis forces.
%

%
%
According to (\ref{eq:tidal_base_flow_O}), the tidal base flow is proportional to the ellipticity of the tidal deformation $\beta$, and may therefore be written as
\begin{equation}
\label{eq:tidal_base_flow_matrix}
\mbf{U}^\Omega = \beta \mathsf{A} (t) \mbf{x}, 
\end{equation}
where $\mathsf{A}(t)$ is a linear operator, harmonic in time with frequency $2\gamma$, acting on the position $\mbf{x}$, representing the rotating strain field depicted in figure \ref{fig:tidal_flow_frame}. 
We recall that $\gamma$ is the difference between the planet's rotation rate and the moon's orbital rate, or equivalently the the rotation rate of the tidal bulge in the rotating frame. 
In the linear, inviscid limit, the equation (\ref{eq:NS_perturb_tidal}) can be recast as 
\begin{equation}
\label{eq:tidal_parametric}
\partial_t \bu + 2 \mbf{e}_z \times \bu~+ \mbf{\nabla} p ~=~ - \beta \left[ \mathsf{A}(t) \bu + \mathsf{A}(t) \mbf{x} \cdot \mbf{\nabla} \bu  \right] ~.
\end{equation}
This equation is very similar to the equation (\ref{eq:parametric_first_order}) governing the length-varying pendulum: eigenmodes of the system (\ie~inertial waves here), represented by the left hand side, are excited by a forcing which is harmonic in time, and depends on the amplitude of the eigenmodes. 
From a qualitative point of view, the periodic stretching of waves by the tidal strain is able to convey energy from tides to waves. 
%

%
%
%
%

%
The situation compared to the length-varying pendulum is here slightly enriched by the existence of an infinite number of oscillators: many pairs of inertial modes can cooperate and resonate with the tidal base flow. 
Let us consider a mode of frequency $\omega_1$: its non-linear interaction with the base flow corresponding to the RHS of (\ref{eq:tidal_parametric}) bears harmonic terms of frequencies  $\pm 2 \gamma +  \omega_1$ that can match a second mode oscillation, hence reinforcing it, provided its frequency $\omega_2$ matches the following resonance condition: 
\begin{equation}
 \vert \omega_1 - \omega_2 \vert = 2 \gamma~.
\end{equation}
Reciprocally, mode 2 then reinforces mode 1. 
There is therefore a coherent effect of the tidal base flow and the two resonant waves in building the parametric resonance. Note that this resonance condition includes the single mode resonance for $\omega_1 = - \omega_2 = \pm \gamma$, which is reminiscent of the length-varying pendulum sub-harmonic resonance.  
Due to the nature of the base flow, this instability has been coined ``elliptical instability''. 
Note also that the resonance condition and the bounded domain of the inertial frequencies implies that the tidally-driven resonance can be excited as long as $ \vert\gamma\vert \leq 2
\vert \Omega \vert$.

\begin{figure}
\centering
\includegraphics[width=\linewidth]{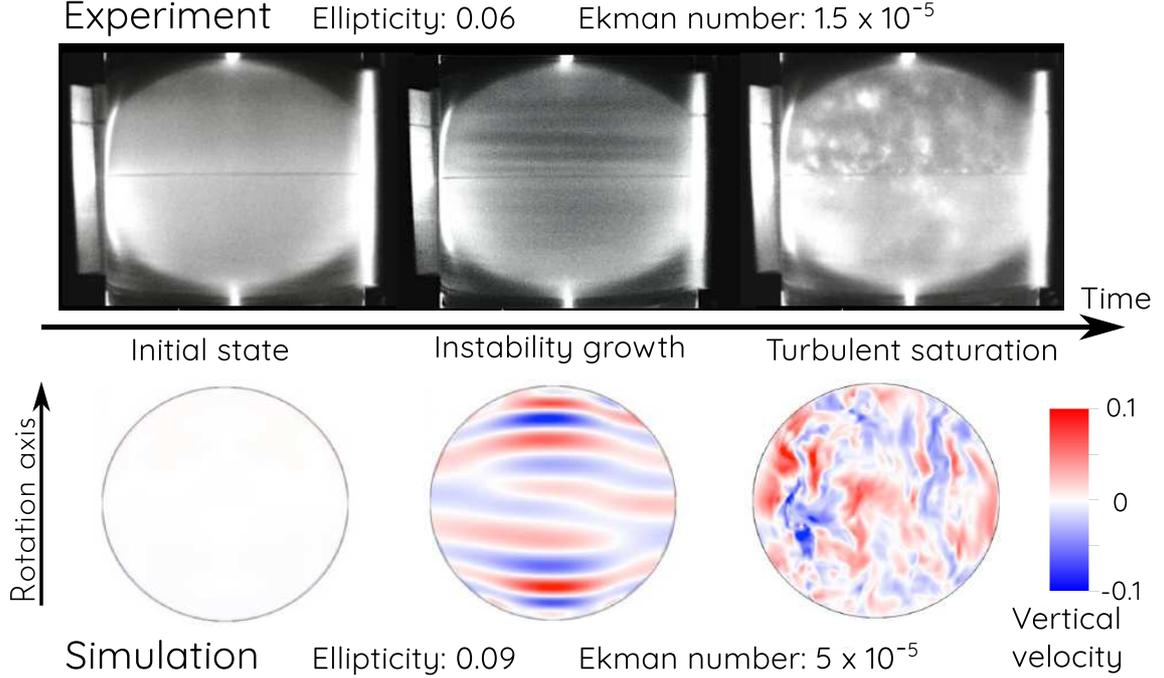}
\caption{Time evolution of the tidally-driven flow inside an ellipsoid shown in a meridional cross-section. \textbf{Top:} experimental visualization using flake-shaped particles materializing shear zones. \textbf{Bottom}: numerical simulation showing the vertical velocity. These pictures are adapted from \cite{grannan_tidally_2017}, Figure 3. }
\label{fig:grannan_2017}
\end{figure}

An illustration of the instability growth in given in figure \ref{fig:grannan_2017}.
In both the experiment and the simulation, the orbital and the spin rate are opposed, that is, $\Omega = -n$, such that $\gamma = 2 \Omega$.
The resonant modes are therefore at the limit of the inertial modes frequency domain. 
These modes are composed of horizontal layers of alternating horizontal velocity. 
They are reminiscent of plane waves in an unbounded domain as, at the frequency $2 \Omega$, the wave vector is purely vertical. 
Lastly, it is interesting to note that even if the ellipticity is quite small in the experiment shown in figure \ref{fig:grannan_2017} ---$\beta = 0.06$---, the flow is fully turbulent at later times, once the instability has reached saturation. 

\subsection{Quantifying the growth rate: a global approach}

In this section, we detail a method to theoretically determine the growth rate of the parametric instability excited by the tidal base flow. 
We look for the long term evolution of the amplitude of the resonant modes, following the general process given in \cite{tilgner_8.07_2007}. 
The perturbation flow $\bu$ can be written as a superposition of two modes with spatial structures $\mbf{\Psi_1}(\mbf{r})$ and $\mbf{\Psi_2}(\mbf{r})$, that is:
\begin{equation}
\label{eq:mode_ansatz}
\bu(\mbf{r},t) ~=~ A_1(t) \mbf{\Psi_1} (\mbf{r}) + A_2(t) \mbf{\Psi_2} (\mbf{r})~,
\end{equation}
where the $A_j (t)$ are the amplitudes of the modes. 
Noting $\Pi_j$ the pressure field associated to the mode $j$, $\mbf{\Psi}_j$ and $\Pi_j$ satisfy the following equation:
\begin{equation}
2 \mbf{e}_z \times \mbf{\Psi}_j + \mbf{\nabla} \Pi_j ~=~  i \omega_j \mbf{\Psi}_j~,
\end{equation}
that is, $\mbf{\Psi}_j, \Pi_j$ are eigenmodes of the linearized, rotating Euler equation. 
Note that, in general, two modes satisfy an orthogonality relation \citep{greenspan_theory_1968}:
\begin{equation}
\label{eq:orthogonality}
\left\langle \mbf{\Psi}_j \bigr\vert \mbf{\Psi}_k \right\rangle ~\equiv~ \frac{1}{\mathcal{V}} \int_{\mathcal{V}} \mbf{\Psi}_j^* \cdot \mbf{\Psi}_k \mathrm{d} V ~=~ \delta_{jk},
\end{equation}
where $\mathcal{V}$ is the domain volume and $\delta$ the Dirac delta function.
Tidal distortion is always a small perturbation of the spherical shape of the container.
The modes inside the slightly distorted container are therefore approximated here by the inertial modes of the sphere for simplicity \cite[but see][for a more complete approach]{vidal_inviscid_2017-1}.
Then we introduce an azimuthal wavenumber $m$ such that the spatial structure of an inertial mode $\mbf{\Psi}$ can be written as 
$$
\mbf{\Psi}(\mbf{r}) ~=~ \mbf{\Phi}(r,z) e^{im\phi}
$$
where $(r,z,\phi)$ are the cylindrical coordinates \citep{greenspan_theory_1968}.
Two structures with different $m$ are orthogonal, and the dot product between two modes can be specified as follows:
\begin{equation}
\label{eq:orthogonality_azimuthal}
\left\langle \mbf{\Psi}_j \bigr\vert \mbf{\Psi}_k \right\rangle ~=~ \delta(m_k-m_j) \frac{2 \pi}{\mathcal{V}}  \int_{r,z} \mbf{\Phi}_j^* \cdot \mbf{\Phi}_k \, r \mathrm{d}r \mathrm{d} z ~=~  \delta(m_k-m_j) \left( \mbf{\Phi}_j \bigr\vert \mbf{\Phi}_k \right) ~.
\end{equation} 
where we have introduced a reduced dot product $\left( \cdot \vert \cdot \right)$ that acts on the radial and vertical structure of the modes. 
We introduce $\mathsf{L}(t)$ the linear operator associated to the RHS of equation (\ref{eq:tidal_parametric}), \ie~the linear operator which couples the modes with the tidal base flow.
In general, for a field $\mbf{w}$, 
\begin{equation}
\mathsf{L}(t) \mbf{w} ~=~  \mathsf{A}(t) \mbf{w} + \mathsf{A}(t) \mbf{x} \cdot \mbf{\nabla} \mbf{w}~.
\end{equation}
With the orthogonality relation (\ref{eq:orthogonality}), the evolution of the amplitudes $A_j$ are inferred from the ansatz (\ref{eq:mode_ansatz}) and the flow equation (\ref{eq:tidal_parametric}). 
\begin{equation}
\label{eq:amplitude_A}
\left\lbrace
\begin{array}{rl}
\dot{A}_1 - i \omega_1 A_1 ~&=~ \beta \left\langle \mbf{\Psi}_1 \bigr\vert L(t) \mbf{u} \right\rangle \\[0.5em]
\dot{A}_2 - i \omega_2 A_2 ~&=~ \beta \left\langle \mbf{\Psi}_2 \bigr\vert L(t) \mbf{u} \right\rangle ~.
\end{array}
\right.
\end{equation}
We may get rid of the fast oscillations of the inertial modes in the preceding equations introducing $a_j (t) = \exp(i \omega_j t) A_j(t)$, similarly to what has been done in section \ref{sec:pendulum}.
The equations (\ref{eq:amplitude_A}) then reads:
\begin{equation}
\label{eq:amplitude_a}
\left\lbrace
\begin{array}{rl}
\dot{a}_1  ~&=~ \beta \left\langle \mbf{\Psi}_1 \bigr\vert L(t) \mbf{u} \right\rangle e^{-i \omega_1 t} \\[0.5em]
\dot{a}_2  ~&=~ \beta \left\langle \mbf{\Psi}_2 \bigr\vert L(t) \mbf{u} \right\rangle e^{-i \omega_2 t} ~.
\end{array}
\right.
\end{equation}
The RHS of (\ref{eq:amplitude_a}) can be specified taking into account the temporal and spatial variations of the modes contained in $\bu$ and in the coupling operator $\mathsf{L}(t)$.
The tidal base flow is transformed into cylindrical coordinate as:
\begin{equation}
\label{eq:cylindrical_base_flow}
\mbf{U}^{\Omega} ~=~ - \beta \gamma r \left[ \sin(2\gamma t + 2 \phi) \mbf{e}_r + \cos(2\gamma t + 2 \phi) \mbf{e}_\phi \right] 
\end{equation}  
showing that it contains the wave numbers $m=2$ and $m=-2$.
$\mathsf{L}$ is therefore  decomposed as:
\begin{equation}
\mathsf{L}(t) ~=~ e^{i (2 \gamma t + 2 \phi)} \mathsf{L}_0 + e^{-i ( 2 \gamma t + 2 \phi)} \mathsf{L}_0^{*}~
\end{equation}
where $\mathsf{L}_0$ is independent of time and $\phi$.
Using the orthogonality relation, the RHS of (\ref{eq:amplitude_a}) is expanded as follows:
\begin{equation}
\label{eq:expansion}
\begin{array}{rl}
\left\langle \mbf{\Psi}_1 \bigr\vert \mathsf{L}(t) \mbf{u} \right\rangle e^{-i \omega_1 t} ~&=~ \left( \mbf{\Phi}_1 \bigr\vert \mathsf{L}_0 \mbf{\Phi}_1  \right) a_1~ \delta(2+m_1-m_1) e^{i ( \omega_1 + 2 \gamma-\omega_1) t} \\[0.5em]
&+ \left( \mbf{\Phi}_1 \bigr\vert \mathsf{L}_0^{*} \mbf{\Phi}_1  \right) a_1 ~ \delta(-2 +m_1 -m_1) e^{i ( \omega_1 - 2 \gamma-\omega_1) t} \\[0.5em]
&+ \left( \mbf{\Phi}_1 \bigr\vert \mathsf{L}_0 \mbf{\Phi}_2  \right) a_2 ~ \delta(2 + m_2 - m_1) e^{i ( \omega_2 + 2 \gamma -\omega_1) t} \\[0.5em]
&+ \left( \mbf{\Phi}_1 \bigr\vert \mathsf{L}_0^{*} \mbf{\Phi}_2  \right) a_2 ~\delta(-2 + m_2 - m_1) e^{i ( \omega_2 - 2 \gamma -\omega_1) t} ~.
\end{array}
\end{equation}
and similarly for $\left\langle \mbf{\Psi}_2 \bigr\vert L(t) \mbf{u} \right\rangle e^{-i \omega_2 t}$.
When the resonance condition is satisfied, \ie~ $\omega_2 - \omega_1 = 2 \gamma$, the coupling between the modes $1$ and $2$ is effective provided the following selection rule applies:
\begin{equation}
\label{eq:selection_rule}
m_2 - m_1 = 2~.
\end{equation}
All the other coupling terms, whose frequencies do not match the resonance condition, then vanish.
A similar derivation for $\left\langle \mbf{\Psi}_2 \bigr\vert \mathsf{L}(t) \mbf{u} \right\rangle e^{-i \omega_2 t}$ allows to prove that the system (\ref{eq:amplitude_a}) reduces to:
\begin{equation}
\label{eq:amplitude_equation_inviscid}
\left\lbrace
\begin{array}{rl}
\dot{a}_1  ~&=~ \beta \left( \mbf{\Phi}_1 \bigr\vert \mathsf{L}_0^{*} \mbf{\Phi}_2  \right)~ a_2 \\[0.5em]
\dot{a}_2  ~&=~ \beta \left( \mbf{\Phi}_2 \bigr\vert \mathsf{L}_0 \mbf{\Phi}_1  \right)~ a_1 ~.
\end{array}
\right.
\end{equation}
The growth rate $\sigma$ is therefore given by the overlap between the tidal base flow and the two modes: 
\begin{equation}
\label{eq:amplitude_growth_rate}
\sigma^2 ~=~ \beta^2  \left( \mbf{\Phi}_1 \bigr\vert \mathsf{L}_0 ^{*}\mbf{\Phi}_2  \right) \left( \mbf{\Phi}_2 \bigr\vert \mathsf{L}_0 \mbf{\Phi}_1  \right).
\end{equation}
The amplitudes of the modes grow provided the overlap integrals have the same sign; the growth rate is then proportional to $\beta$. 
This is similar to the length-varying pendulum for which the growth rate was found to be proportional to the amplitude of the perturbation $\eta$. 
Computing the growth rate is in this case rather difficult as it requires computing the overlap integrals between the modes and the tidal forcing. 
This is in general non-trivial: although the inertial modes in a sphere, or even a spheroid, are known, there is no analytic formula in the generic case of tri-axial ellipsoids. 
Computation of the overlap integrals therefore requires numerical solving of the eigenvalue problem of inertial modes, as done for instance in \cite{vidal_inviscid_2017-1}.
Lastly, the amplitude equations (\ref{eq:amplitude_equation_inviscid}) may be refined accounting for the viscous damping of the modes. 
In the planetary limit of small Ekman number $E$, viscous dissipation is dominated by friction inside Ekman boundary layers, and yields a correction $\mathcal{O}(\sqrt{E})$; as shown by \cite{le_bars_tidal_2010}, the correction to the growth rate is then $K \sqrt{E}$ with $K$ a constant typically between 1 and 10 \cite[but see also][for a discussion on the possible importance of bulk dissipation]{lemasquerier_d._librationdriven_2017}. 

\subsection{Quantifying the growth rate: short wavelength approximation}
Although the derivation of the growth rate in the preceding paragraph is rather straightforward, it is quite difficult to extract quantitative information in complex geometries such as tri-axial ellipsoids. 
This approach requires knowing \textit{a priori} the spatial structure of the inertial modes that must be computed numerically.
Another approach that has proven efficient in past studies to make quantitative prediction (see e.g. \cite{kerswell_elliptical_2002} and references therein) consists in assuming a scale separation between short wavenumber resonant modes and the large scale tidal flow. 
This approach, known as the Wentzel-Kramers-Brillouin (WKB) analysis, assumes that the perturbations take the form of a plane wave packet around a point that is advected by the base flow. 
The wave packet is affected by tidal distortion as it moves along with the Lagrangian point. 
This theoretical framework, which resembles the process to infer classical optics from light wave propagation, was formally introduced in the context of hydrodynamic instabilities by \cite{lifschitz_local_1991}.
It is particularly suitable for the study of parametric instability of waves interacting with a base flow. 

It was applied to the present case of the tidally driven elliptical instability in rotating flows by \cite{le_dizes_three-dimensional_2000}. 
The WKB method allows to retrieve that the short wavelength growing perturbation corresponds to the superposition of two contrapropagating inertial waves of frequency $\pm \gamma$ with an amplitude growth rate 
\begin{equation}
\sigma =  \frac{\beta \gamma}{16} \left( 2+\gamma \right)^2 ~.
\end{equation} 
Although this growth rate describes short wavelength perturbations under the form of inviscid plane waves in an infinite domain, it accounts very well for the growth of inertial modes in enclosed containers in the weak tidal distortion and low dissipation regime, \ie~the regime that is relevant for geophysics \citep{le_bars_tidal_2010}. A small correction due to boundary friction must then be considered, that is:
\begin{equation}
\sigma_v ~=~ \sigma - K \sqrt{E}
\end{equation}
with $K$ between 1 and 10 typically, as explained in the preceding section. 
\subsection{The elliptical instability in planetary cores}
The preceding theoretical results have been used in past studies to evaluate the actual relevance of the tidally driven elliptical instability in natural systems, for instance by \cite{cebron_elliptical_2012}. 
In the case of the Earth, the tidal distortion of the Moon ---which induces a tidal bulge of ellipticity $\beta \sim 10^{-7}$--- is close but \textit{a priori} not sufficient to overcome the viscous damping of resonant modes. 
Nevertheless, the study \cite{cebron_elliptical_2012} also considered the early history of the Earth at times where the Moon was closer to its parent planet. 
Assuming the Earth-Moon distance is reduced by a factor 2, the core of the Earth becomes unstable to the elliptical instability.
Mechanical forcings therefore provide an interesting alternative to drive magnetic field before the crystallization of the inner core. 
Remember that the Earth is known to be surrounded by a magnetic field since at least 3.5 Gy \citep{tarduno_geodynamo_2010} while the Earth's core only started to form around 1 Gy ago \citep{labrosse_thermal_2015}.
Following the same process as for the tidally-driven instability, librations have also been shown to drive parametric sub-harmonic instability of inertial waves \citep{kerswell_tidal_1998, cebron2012libration, grannan_experimental_2014, cebron_libration-driven_2014}.
In particular, \cite{kerswell_tidal_1998} have demonstrated that Io's core in unstable, thus providing a possible explanation to the magnetic field measured by the Galileo probe around the Jovian satellite \citep{kivelson_ios_1996}.
\cite{cebron_elliptical_2012} have extended the preceding study to show that Europa's core may also be unstable.
Note that Ganymede's core remains below the threshold of instability despite being surrounded by a magnetic field. 
Finally, precession is similarly capable of driving parametric sub-harmonic instability of inertial waves, but there the coupling flow still remains controversial \cite[see e.g.][]{kerswell_instability_1993, lin2015shear}. 
This precession driven instability could be of importance for instance for the magnetic field of the early Earth \citep{andrault2016deep} and Moon \citep{dwyer_long-lived_2011}.

\section{An overview of some ongoing research}

\subsection{Present tools for investigating mechanical forcings and instabilities in the laboratory}

\begin{figure}
\centering
\includegraphics[width=\linewidth]{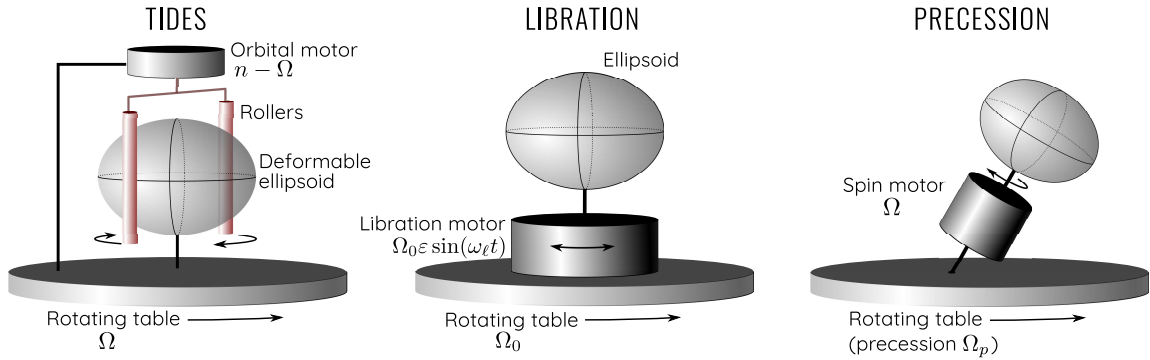}
\caption{A sketch of the three experimental setups built to study flows driven by tides, libration and precession. The rotation rates of the different motors are associated with geophysical forcings. The different rotation rates and amplitudes are defined in section 1. }
\label{fig:setups}
\end{figure}

The investigation of instabilities forced by tides, libration and precession have greatly benefited from experimental studies since the early work of  \cite{malkus_precessional_1963,malkus_precession_1968,malkus_experimental_1989}. 
The typical setups used to study flows driven by tides, libration and precession are given in figure \ref{fig:setups}. 
The interested reader should refer to the review of \cite{le_bars_flows_2015}, or research articles such as \cite{grannan_tidally_2017} and \cite{noir_experimental_2001}. 
Tidal forcing is the most difficult as it requires using a deformable ellipsoid (usually made of silicon); the motion of the tidal deformation is imposed by rollers pressing on the ellipsoid.
Today, modern metrology techniques like high-speed, time-resolved particle imaging velocimetry allow describing quantitatively the turbulence in asymptotic regimes of low dissipation and low forcing amplitude bearing some resemblance with planetary ones \citep{grannan_tidally_2017}.

Additionally, high performance computing now allows reaching fully turbulent cases and considering the fully coupled magneto-hydrodynamics problem for dynamo action. Two main strategies have been developed: either consider so-called ``local'' numerical methods (like finite elements, finite volumes, or spectral elements) to simulate the flow in the tri-axial ellipsoidal geometry relevant for planetary applications \cite[e.g][]{cebron2010systematic, ernst2013, favier_generation_2015}; or use highly-efficient numerical methods in simplified geometries (like a sphere or a triply periodic box), then imposing the mechanical forcing as a background force \cite[e.g.][]{barker_non-linear_2013, Cebron:AJL2014, le_reun_inertial_2017}.

All methods have pro and cons; only their combination --together with real data analysis-- has allowed significant progress towards a better understanding of those flows and their planetary consequences. 
Some of these advances are presented below.

\subsection{The saturation of the instability}

Since the seminal studies by Malkus in the 1960s, many investigations have focused on showing the relevance of mechanical forcing for actual planetary cores.
However, only a few works have been devoted to the saturation of the instability, but it is vitally important for tidal dissipation, the orbital evolution of planets and stars, and dynamo action. 
In the various experimental and numerical investigations of the saturation flow, a large diversity of behaviors have been observed, depending on the geometry and the control parameters ---be it the tidal distortion, the Ekman number or the libration amplitude and frequency.
In the saturation flow, the emergence of geostrophic vortices, \ie~flows that are invariant along the rotation axis and that have a slow dynamics,\footnote{They correspond to the $\omega \rightarrow 0$ limit of inertial waves.} seems to play an important role.
As they grow, inertial modes feed a geostrophic flow, probably via a secondary instability as shown by \cite{kerswell_secondary_1999}, which might become of similar amplitude as the excitation flow and lead to the disruption of the instability \citep{barker_non-linear_2013}.
The flow then decays viscously before waves can again be excited and engage into a new cycle of growth and decay. 
Some experiments however, such as those carried out by \cite{grannan_experimental_2014}, have shown that the saturation flow can also be fully turbulent and reach a statistically steady state.
Further simulations carried out by \cite{favier_generation_2015} have even shown that the unstable inertial modes are able to excite \textit{a priori} stable modes, \ie~outside the initial resonance condition, via three modes resonant interactions, a mechanism called ``triadic resonance'' ---see for instance the work of \cite{bordes_experimental_2012}.
This discrepancy was somewhat clarified by \cite{le_reun_inertial_2017}.
This study used the method introduced by \cite{barker_non-linear_2013} to study the tidally-driven elliptical instability: instead of simulating the whole planetary interior with complex ellipsoidal outer boundary, only a small parcel is modeled. 
This parcel can be assumed to be cubic, with periodic boundary conditions, and the tidal excitation is imposed as a background flow. 
Such a technique allows to study with great details the saturation flow in the regime of low forcing (small ellipticity) and low dissipation (small Ekman number) which is relevant for geophysics.
With an artificial and adjustable friction specific to the geostrophic component of the flow, the saturation continuously transitions from a geostrophic-dominated turbulence to a wave-dominated turbulence.
In the later case, it was even shown that the resulting turbulent flow is a superposition of a swarm of inertial waves with different frequencies, which are all in resonant interaction, a state  known as an inertial wave turbulence \citep{galtier2003weak,yarom2014experimental}. This regime is expected in the asymptotic limit (low dissipation and low forcing) relevant for planetary flows.
\subsection{Dynamo driven by mechanical forcing}
One of the most striking features of the past few years was to show that the mechanically driven flows are able to drive dynamo action in planetary cores. 
The few published studies include laminar dynamos from the precession base flow \citep{ernst2013}, laminar dynamos for tidal instability with \textit{ad hoc} bulk forcing in a spherical domain \citep{Cebron:AJL2014,vidal2017magnetic}, laminar dynamos in a spheroidal domain for precession and libration instabilities \citep{Wu:GAFD2009,Wu:GAFD2013}, and turbulent dynamos in a spherical domain for precession \citep{Tilgner:POF2005,tilgner2007kinematic,kida2011turbulent,Lin:POF2016}.
\cite{reddy_k._sandeep_turbulent_2018} have carried out simulations of libration, precession and tidally driven instabilities in tri-axial ellipsoids leading to sustained turbulence, which is the relevant configuration for planetary applications. 
They showed that such a flow is able to drive amplification of the magnetic field in the three cases. 
However, the feedback of the Lorentz force was not taken into account in this work ---\ie~the authors addressed kinematic dynamos only.
The full problem involving sustained turbulence, magnetic field and the Lorentz force in tri-axial ellipsoids, remains to be explored. 
In particular, the different features of the magnetic field induced by an inertial wave turbulence or a geostrophic turbulence, as discussed in the preceding paragraph, have only been sparsely investigated \citep{moffatt_dynamo_1970,davidson_dynamics_2014}.

\subsection{Instabilities in presence of an inner core}

The whole set of results discussed above were all obtained in the simple context of fully liquid cores. 
We know that the secular cooling of planets sometimes leads to the progressive crystallization of a solid inner core (like in the Earth).
Considering the relevance of mechanical forcing in a shell with inner and outer boundaries is more complicated as the inertial modes become singular; that is, they are locally discontinuous on sheets in the inviscid limit ---see for instance discussions in \cite{rieutord_inertial_2001}.
From a mathematical point of view, the existence of such singular solutions is due to the hyperbolic nature of the Poincar\'e equation describing wave propagation, for which conditions are applied at the boundaries of the domain, which implies ill-posedness.
A few studies have been devoted to tidally-driven instabilities with an inner core, mostly in the case of tidal and libration forcings \citep{seyed-mahmoud_elliptical_2004,lacaze_elliptical_2005-1,lemasquerier_d._librationdriven_2017}.
In these experiments, the elliptical instability mechanism, \ie~ the parametric excitation of inertial waves by the base flow, still holds.
Nevertheless, the turbulent saturation flow happens to be heterogeneous with different features inside and outside the cylinder parallel to the axis of rotation and tangent to the inner core \citep{lemasquerier_d._librationdriven_2017}.

\subsection{The mysterious magnetic field of the early Moon}

We conclude with a section on how the elliptical instability explains the early history of the Moon, which used to have a magnetic field, even when its core was not convecting.
M

\subsubsection{The early magnetic field of the Moon}

\begin{figure}
\centering
\includegraphics[width=0.8\linewidth]{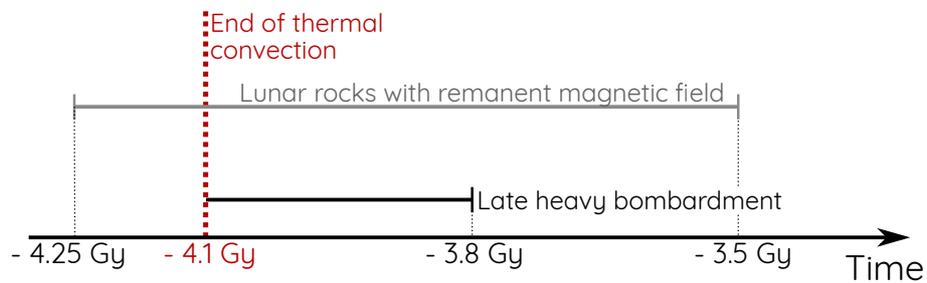}
\caption{
Time line of the events affecting the early Moon with focus on the magnetic era, from \cite{weiss_lunar_2014} and \cite{le_bars_impact-driven_2011}.}
\label{fig:moon_magnetic_field}
\end{figure}

\begin{figure}
\centering
\includegraphics[width=\linewidth]{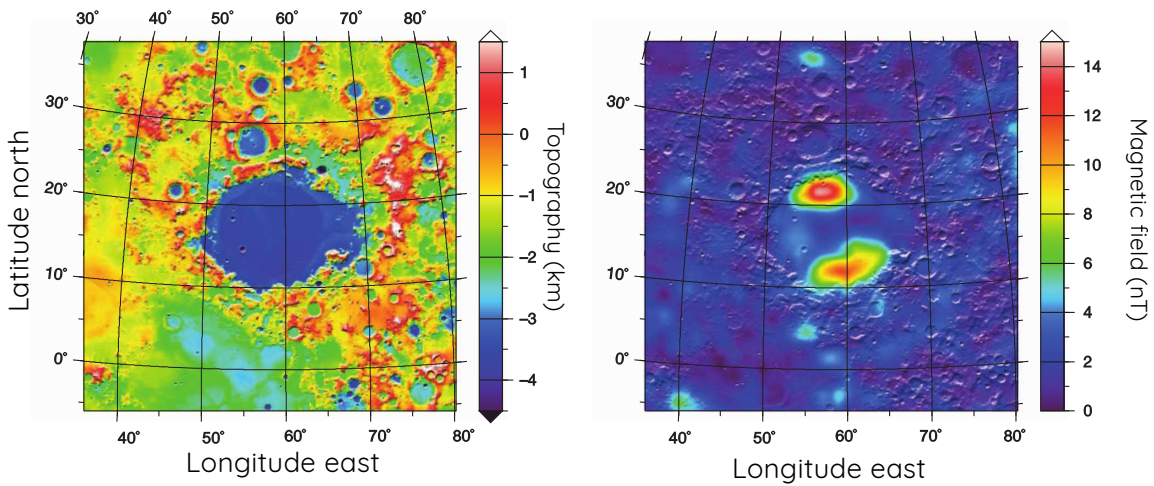}
\caption{ Surface topography and corresponding strength of the magnetic field of the Crisium impact basin, reproduced after \cite{le_bars_impact-driven_2011}, Figure 1. The largest values of the magnetic field are restricted to the impact basin where a thick melt event occurred. 
}
\label{fig:lebars2011_impact}
\end{figure}

From measurements of the remanant magnetization of the surface of the Moon, determined on samples collected during the Apollo missions, planetologists show that the Moon used to be surrounded by a strong magnetic field during 700 to 800 million years, starting around 4 billion years ago \citep{weiss_lunar_2014} --- see the schematic in figure \ref{fig:moon_magnetic_field}.
According to \cite{garrick-bethell_early_2009}, the minimum intensity of the magnetic field, inferred from the above-mentioned measurements, was around $1~\mu$T. 
This finding seems puzzling as thermal evolution models of the Moon state that thermal convection is not possible in its core after -4 Gy. 
Satellite data show a strong correlation between the location of the largest values of the magnetic field strength and large impact basins where the collision is thought to have induced a thick melting of the Moon's crust and mantle ---see figure \ref{fig:lebars2011_impact}. 
The time line is also striking as the age of the basins where the strongest remanant magnetization are observed correspond to the late heavy bombardment, an era when several impacts of large bodies occurred approximately 4.1 to 3.8 billion years ago. 
These data therefore show a strong correlation between large meteoric collisions and the existence of a magnetic field. 

\subsubsection{Desynchronisation dynamo in the Moon}

Before an impact, the rotation rate of the Moon is synchronized with its revolution around the Earth. 
In their work, \cite{le_bars_impact-driven_2011} pointed out that large impacts are able to change the rotation rate of the Moon, and more precisely the rotation rate of its mantle.
The fluid outer core of the Moon is coupled to the mantle via viscous drag at the boundary.
Therefore, after an impact, the core and the mantle rotate at different rates. 
In addition, the Moon is known to have a frozen, or fossil tidal bulge: it has a deformation that is reminiscent of a tidal bulge, but is not forced by any actual gravitational interaction (i.e. it accounts for a rise of 200 m while the present tidal bulge height stands around 10 m). This deformation --- or a largest tidal bulge --- was thus present in the past.
Note that such a frozen distortion is not in hydrostatic equilibrium; the gravitational field of the Moon is not strong enough to force the relaxation of this imbalance.

\begin{figure}
\centering
\includegraphics[width=0.6\linewidth]{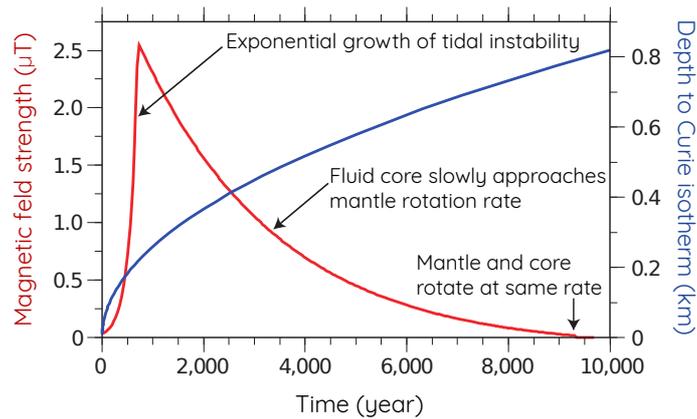}
\caption{Time evolution of the magnetic field strength and the depth of the Curie isotherm below which rocks acquire a remanant magnetic field. Figure adapted from \cite{le_bars_impact-driven_2011}, figure 3. 
}
\label{fig:lebars2011_dynamo}
\end{figure}

After an impact, the core thus sees a deformation that rotates at the mantle's rate, a configuration that exactly matches the tidal forcing which is known to excite parametric instability of inertial waves.
\cite{le_bars_impact-driven_2011} proved that desynchronization by large impacts is sufficient to drive tidally-driven elliptical instability, and to potentially induce a dynamo.
Note that the Moon was closer to the Earth at that time, at about 45 Earth's radii, while it is at 60 Earth's radii today. 
As a consequence, the Moon's orbital rate was larger, thus lowering the Ekman number and the dissipation, which facilitated the growth of the instability. 
They proposed that the instability takes around 1000 years to develop and saturate.
The turbulence driven by the non-synchronous rotation drives enhanced tidal dissipation which tends to resynchronize the spin and the orbit of the Moon, a process that takes around 10~000 years. 
During that transition period, the magnetic field may have reached up to $2.5 ~\mu$T, as indicated in figure \ref{fig:lebars2011_dynamo}.
The depth of the Curie isotherm, below which rocks acquire remanant magnetization, indicates that the surface of the impact basin could record the magnetic field at its peak ---see figure \ref{fig:lebars2011_dynamo}.
For a less intense impact, free libration, instead of full desynchronization, might have similarly led to elliptical instability, turbulence, dynamo and recording at the surface while cooling and slowing down.

Note finally that an analog scenario was proposed at the same time by \cite{dwyer_long-lived_2011}, based on the more intense precession forcing that was present at that time. 
It is worth considering that the different scenarios are not exclusive, but might be combined to explain the whole story of the Moon magnetic field evolution. 

\section{Conclusion --- Beyond the elliptical instability in planetary cores}

\begin{figure}
\includegraphics[width=\linewidth]{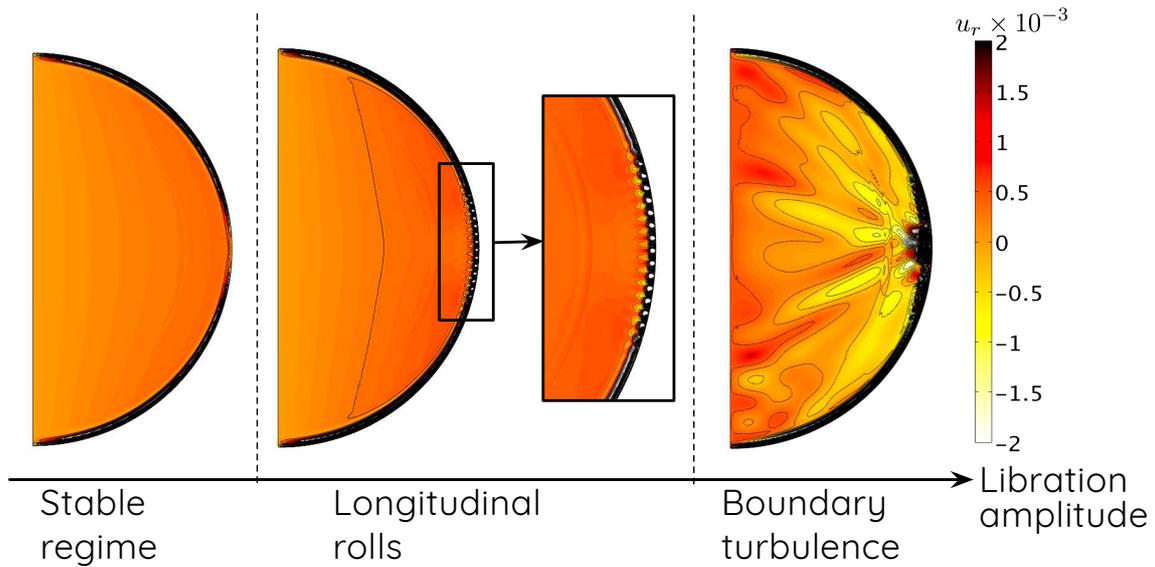}
\caption{Illustration of the destabilization of the equatorial boundary layer caused by librations. The color scale represents the radial velocity. This figure is adapted from \cite{sauret_spontaneous_2013} Figure 2.}
\label{fig:sauret2013}
\end{figure}

In this chapter, we have presented how gravitational interactions are able to create complex, and even turbulent flows, inside planetary cores.
We have particularly focused on a specific class of instability related to parametric sub-harmonic resonance, drawing an analogy between rotating fluid cavities and a length-varying pendulum. 
Such a resonance, also called ``elliptical instability'', builds on the coupling between two inertial waves and the oscillating strain field induced by tidal interactions.
Despite recent advances that have been presented above, many challenges remain to describe the turbulent saturation flows and the induced magnetic field.
Also, while tides distortion, libration, precession and convection have up to now been mainly studied separately, they might be simultaneously present in planetary cores, possibly leading to complex non-linear interactions that remain to be explored.

A concluding remark: parametric resonance is not the only class of instabilities that can be driven by tides inside planetary cores. 
For instance, libration leads to centrifugal instability of the equatorial boundary layer \citep{calkins_axisymmetric_2010,favier_non-linear_2014}. 
As shown numerically by \cite{sauret_spontaneous_2013}, this excited motion in the boundary layer radiates inertial waves in the whole interior ---see figure \ref{fig:sauret2013}.
Lastly, the excitation of inertial modes inside a fluid cavity drives, via interactions with and within the boundary layer, zonal winds, \ie~axisymmetric geostrophic flows, associated with intense shear that may be unstable \citep{sauret_experimental_2010,sauret_tide-driven_2014}.
All these different excited flows pile up in planetary cores and their respective contributions remain to be evaluated. 
In any case, we hope to have convinced the reader that tidal interactions lead to complexity in the behavior of fluid interiors, and should not be dismissed for being ``small'' perturbations. 

\newpage

%


\setlength{\bibsep}{3pt plus 0.3ex}

%

\end{document}